%% file: main.tex
\documentclass{article}

\usepackage{arxiv}

\usepackage[utf8]{inputenc} 
\usepackage[T1]{fontenc}    
\usepackage{hyperref}       
\usepackage{url}            
\usepackage{booktabs}       
\usepackage{amsfonts}       
\usepackage{nicefrac}       
\usepackage{microtype}      
\usepackage{lipsum}

\usepackage{amsmath}
\usepackage{color}
\usepackage{graphicx}
\usepackage{listings}
\usepackage{natbib}
\usepackage{subcaption}
\usepackage[normalem]{ulem}

\usepackage{siunitx}
\title{%
    SHIFT: A Highly Realistic Financial Market Simulation Platform \\
}

\author{%
  Thiago W. Alves$^{\ast}$ \\
  School of Business \\
  Stevens Institute of Technology \\
  Hoboken, NJ 07310 \\
  \texttt{twinkle1@stevens.edu} \\
  \And
  Ionu\c{t} Florescu \\
  School of Business \\
  Stevens Institute of Technology \\
  Hoboken, NJ 07310 \\
  \texttt{ifloresc@stevens.edu} \\
  \And
  George Calhoun \\
  School of Business \\
  Stevens Institute of Technology \\
  Hoboken, NJ 07310 \\
  \texttt{gcalhoun@stevens.edu} \\
  \And
  Drago\c{s} Bozdog \\
  School of Business \\
  Stevens Institute of Technology \\
  Hoboken, NJ 07310 \\
  \texttt{dbozdog@stevens.edu} \\
}

\begin{document}
\maketitle

{
\small
\textbf{Funding Information} \\
CME Group Foundation Research Grants for the SHIFT Project (2016, 2017) \\
$^{\ast}$ CNPq Science Without Borders Grant/Award Number: 200989/2014-6
}

\vskip 0.4in

\begin{abstract}
This paper presents a new financial market simulator that may be used as a tool in both industry and academia for research in market microstructure. It allows multiple automated traders and/or researchers to simultaneously connect to an exchange-like environment, where they are able to asynchronously trade several financial assets at the same time. In its current iteration, this order-driven market implements the basic rules of U.S. equity markets, supporting both market and limit orders, and executing them in a first-in-first-out fashion. We overview the system architecture and we present possible use cases. We demonstrate how a set of automated agents is capable of producing a price process with characteristics similar to the statistics of real price from financial markets. Finally, we detail a market stress scenario and we draw, what we believe to be, interesting conclusions about crash events.
\end{abstract}


\input{introduction}
\input{description}
\input{replay}
\input{simulation}
\input{conclusion}

\section*{Acknowledgements}
This project would not be possible without the contribution of several Stevens Institute of Technology students and researchers who participated in different stages of its development: Chen Liu, Gaojie Li, Han Zheng, Hanrun Li, Isaac Cohen, Jian Zhao, Jiaxu Duan, Jinyu Zeng, Lalita Gajbe, Meng Zhi, Runxi Ding, Shaoyong Tang, Shiwei Zeng, Shuoyu Mao, Waris Bantherngpaesach, Weipu Xu, Xiaojian Zhu, Xiaoshuai Luo, Xuan Luo, Xuming Bing, Yang Liu, Yongxin Feng, Yuan Tian, Yuewei Mao, Zhanyu Tan, Zhenjiu Dai, Ziwen Ye.

\bibliographystyle{chicago}
\bibliography{references}

\appendix
\input{api}

\end{document}

%% file: introduction.tex
\section{Introduction}
\label{sec:introduction}

A recent Congressional Research Service report on \textit{high frequency trading} \citep{miller_high_2016} estimates that it accounts for $55\%$ of the U.S. equity market and $40\%$ of European equity markets. Many studies have been done on the advantages and disadvantages such group of traders pose to the health of financial markets. Some are discussed in \cite{ye_extracting_2019}. High frequency trading, however, is just a subset of larger \textit{algorithmic trading}.

There are many possible interpretations to what algorithmic trading actually means. In general, it refers to advanced mathematical models used in either automatic trading strategies or optimal order execution algorithms, with little to no human interaction. \cite{kissell_science_2013} estimates that algorithmic trading as a whole accounted for $85\%$ of market volume in 2012. A report from Research and Markets\footnote{Algorithmic Trading Market by Trading Type (FOREX, Stock Markets, ETF, Bonds, Cryptocurrencies), Component (Solutions and Services), Deployment Mode (Cloud and On-premises), Enterprise Size, and Region - Global Forecast to 2024.} estimates the global algorithmic trading market size to grow from $11.1$ billion U.S. dollars in 2019 to $18.8$ billion U.S. dollars by 2024. In fact, according to JPMorgan analysts, only $10\%$ of 2017's stock market trading volume was performed by fundamental value traders. Among other possible effects, this high activity of algorithmic-controlled trading may cause sell-off episodes when machines act immediately after data releases, without the proper analysis a human would do.

In order to try and mitigate the effects of ill-constructed algorithms, regulations such as Regulation Automated Trading (``Reg AT'') by the Commodity Futures Trading Commission (``CFTC'') \citep{cftc_regulation_2015} of the United States (one of the main regulators for derivatives markets in the U.S.) have been proposed. In general terms, Reg AT recognizes the urgent need to adapt financial market regulations to the current business models under which exchanges and most traders are operating today. Specifically, today's trades are based on high-speed automated market processes for all segments of typical financial transactions, from order placement and cancellation, to the operation of matching engines for connecting and clearing bids and offers, to post-trade processing and data reporting. The CFTC points out in its extensive and ambitious rule-making effort \citep{cftc_regulation_2015} that most of its supervisory policies assume a world in which trades are executed ``by hand'' -- with extensive human intermediation, and at ``human speed'' -- whereas the technology in the market today operates at ``machine speed'' with latency as low as a few hundred microseconds. The broad charter of Reg AT calls for a rethinking and revamping of market regulation to bring the framework up to date with modern technology.

Reg AT specifically calls for development of capability to \textit{test} all forms of automated trading or trade-support systems that interface directly with financial markets before they are introduced into a real exchange environment. This capability should operate in a controlled, off-line \textit{test-bed} environment -- but one that is realistic enough to allow reasonable assessment of the likely impact and risk of operating those systems in a ``live'' market.

We believe such a test-bed would allow exchange operators to explore the consequences of possible rule changes, new order type offerings, or anti-spoofing measures. The system would potentially have value for private sector participants who wish to test the effectiveness of algorithmic trading systems. Moreover, researchers proposing specific changes to the way markets operate, e.g. \cite{budish_high-frequency_2015}, could benefit from such platform. Most of the current research involving policy changes are either based on a theoretical framework, with no empirical evidence that rules would properly work in practice, or they study the consequences of a rule change implementation on a particular exchange, months after the fact \citep{jorgensen_throttling_2018}.

This technical capability does not yet exist, and Reg AT is vague as to the requirements such a system would have to meet. The models developed thus far are primarily based on agent-based simulations. These existing models are generally limited -- often based on a agents trading a single instrument, with simulated low-frequency data and highly artificial trading rules.

The simulator described in this paper was constructed with the goal of creating a test environment as close to reality as possible. We replicated all the basic characteristics of a financial market exchange and tried to expand on what is presented in agent-based models literature. The final result is a system that is very versatile and can be applied to different scenarios in education and research.


\subsection{Reviewing Agent-Based Models in Finance}
\label{subsec:literature}

The Santa Fe Artificial Stock Market \citep{palmer_artificial_1999} is one of the most cited agent-based systems applied to finance. With the development work taking place in the 90's, the system models a market with one risk-free bond and a single stock traded by agents, which follow a set of pre-defined basic rules. The system is now viewed as ground breaking since it is the first one that models traders and measures the result of their interaction i.e., the equity behavior. 

Another, more developed market simulator is presented in \cite{jacobs_financial_2004}. The authors point to the fact that although asynchronous-time, discrete-event simulations are commonly used to model complex systems, they are rarely used to model financial markets. The system is a multi-asset trading environment with asynchronous events. The authors describe these asynchronous events as the agent states are updated at different times (not all agents are updated at every turn). Outstanding buy and sell orders remain in a book, and simulation sessions last several (virtual) days, with trading events happening throughout the day. The agents are mean-variance portfolio holders that trade at most once a day.

The next references are important as they detail a system used to study multiple aspects of financial markets. The Genoa Artificial Stock Market (GASM) \citep{raberto_agent-based_2001} is a market simulator that serves as basis for multiple research papers published since 2000. The authors set to build a simple market structure that would be able to reproduce some stylized facts, such as \textit{volatility clustering} and \textit{heavy tails}, observable in the distribution of real data returns. There is only one risky asset in the market and agents send random limit orders at every simulation step, based on a finite amount of cash and the current realized volatility. Price is formed by the intersection of the demand and supply curves (since the system does not implement a limit order book). \cite{raberto_traders_2003} increments the simulation with different agent strategies, and compares their performance by looking at their wealth evolution. \cite{cincotti_who_2003} adds multi-asset support to the simulator. Each agent is now holding a portfolio, with no short positions allowed. Most of the agents act in a completely random fashion, but the paper explores the application of three different trading strategies (mean-variance, mean-reversion and relative chartist strategy) acting in the resulting market. Both \cite{ponta_information-based_2011} and \cite{ponta_traders_2018} explore information exchange networks among traders in variations of this multi-asset market. In \cite{raberto_modeling_2005} and \cite{ponta_statistical_2012}, the single-asset model from \cite{raberto_agent-based_2001} is extended to use a limit order book as its pricing mechanism. To accommodate the simulation to this new pricing mechanism (with time-price priority), one single agent is chosen at random at every time step to perform an action.

\cite{jacob_leal_rock_2016} proposes a model designed to study the interaction between low frequency traders and high frequency traders (HFT) on a single-asset market. Slow (low frequency) traders submit orders every $\theta$ turns, with each agent having a different $\theta$, based on either a fundamentalist or a chartist strategy. The orders sent by low frequency traders are sent ahead of other agents at every simulation step. High speed traders act every time they see a profit opportunity, employing directional strategies. They submit their orders after the submissions from low frequency traders are completed. The idea being that they are fast enough to exploit the information generated by slow traders. The authors conclude that their approach is able to reproduce main stylized facts from current financial markets. We review this paper as one of the few examples of agent based models that is attempting to model the low frequency/high frequency interaction. Please note that the framework is a turn based simulator similar with the traditional ones described above.  

A thorough review of such agent-based simulation studies is presented in \cite{alsulaiman_bounded_2015}. In general, authors set to solve a research problem. They adapt the most suitable agent-based simulator to answer the research problem studied. They generally focus on a few sets of features that are affecting the research problem. Once the problem has been answered and a new problem appears likely the old simulator needs to be redone. The GASM model mentioned above is symptomatic in this aspect as every paper added a new layer of complexity to be able to answer a new problem in effect evolving the system toward a more realistic one. 

In 2014, when we started the development of the system described in this paper, we wanted a ``as close to reality as possible'' replica of a real market exchange. To this end we created and replicated a real market. This task was extremely complicated and in fact we rebuilt the system from scratch four times to be the completely expandable system we have today. We believe the resulting SHIFT\footnote{SHIFT =  ``\textbf{S}tevens \textbf{Hi}gh \textbf{F}requency \textbf{T}rading Market Simulation System''} system described in Section \ref{sec:description} behaves as close to a real market as possible in a research environment. In fact, we can trade any real standardized asset in the SHIFT system. We shall discuss this in Sections \ref{sec:replay} and \ref{sec:simulation}.

When comparing SHIFT with the existing agent-based models we found three features that are \textit{all} present in our system and which we think are crucial to replicate how markets operate today. The artificial markets in existing literature may contain one or at most two of these features. These features in order of importance are: \textbf{real pricing mechanism}, \textbf{distributed asynchronous}, and \textbf{multi-asset}. 

\paragraph{Real pricing mechanism.} Most exchanges today are \textit{order driven}, while the rest are \textit{quote driven}. Both types of exchanges as well as exchange participants need to keep track of supply and demand as these are the main drivers of market microstructure. \cite{alsulaiman_bounded_2015} cites only four studies which use the limit order book as the pricing mechanism and no quote driven markets.

\paragraph{Distributed asynchronous market.} The majority of financial market simulators in the literature employ some type of ``turn-based system''. Even if the agents do not ``play'' at every turn, e.g. they perform an action every $\Delta t$, most of the times there is a notion of action taken at step $t = 1...T$ and a central unit dictating the order of agent turns.\footnote{Some studies randomize the order of action of the agents at every turn \citep{fricke_effects_2015}.} Despite some serious attempts to introduce a real batch auction exchange, operating in discrete time \citep{budish_high-frequency_2015}, all exchanges today operate in real time. Therefore, we believe that having a \textit{distributed asynchronous} system where clients may be all over the world dealing with real latency as well as a market exchange operator processing the orders in the order that they arrive is crucial to be able to simulate a high frequency trading environment. \cite{jacobs_financial_2004} goes in this direction with its implementation of asynchronous events, but even though events are rendered randomly or are caused by other events, the central unit controlling the simulation knows what the next event will be.

In our system, agents perform actions whenever they want to, and the central unit is constantly listening for incoming messages, with no control over when they are sent and by whom. In a turn based system, high frequency traders are commonly simulated using a smaller $\Delta t$, and thus the orders from the low frequency traders never arrive before their orders. However, in a real system low frequency orders operating on outdated information may arrive earlier at the exchange and by chance predate the HFT orders.

\paragraph{Multi-asset market.} Most of the academic literature employing agent-based simulators is using a single risky asset and a risk free asset. This one-traded-asset model is certainly the basis of any simulation, and many interesting conclusions may be derived. However, allowing agents to trade multiple assets can potentially recreate the highly correlated markets we are experiencing today. We note \cite{cincotti_who_2003} as early work using a portfolio of traded assets. Furthermore, the ability to trade an ETF (i.e., a basket of stocks), as well as the ETF's components allows us to study complex events such as the May $2010$ Flash Crash \citep{kirilenko_flash_2017,paddrik_agent_2012}.

We would like to make a special mention of the Penn-Lehman Automated Trading Project \citep{kearns_penn-lehman_2003}, developed by the Computer Science Department at Pennsylvania State University in partnership with Lehman Brothers. The concept was similar with our system,\footnote{According to \url{http://www.cis.upenn.edu/~mkearns/projects/plat.html}, the project is not active anymore.} but it was limited to single-asset trading. Further, it required a constant feed of real market data (either historical or live every $3$ seconds) to operate (probably to provide liquidity to its users). Although the system was used to organize algorithmic trading competitions, we were not able to find any evidence of agents trading against each other or being capable to move the market through their trades.


\subsection{Focus of this Paper}
\label{subsec:focus}

This paper is focused on defining a realistic test bed capability, extending the agent-based approach to encompass a much higher degree of realism in a rich market environment capable of dealing with:
\begin{itemize}
	\item Large numbers of agents trading large numbers of assets.
    \item Realistic and robust trading strategies.
    \item Real-time, high-frequency market pricing and limit order book data.
    \item The ability to observe interactions between multiple agents (traders), employing potentially overlapping and competing strategies, to enable the study of realistic market events such as crowded trades and liquidity crises.
    \item The ability to test under realistic conditions the effects of regulatory measures, either imposed by a central regulator (e.g., the CFTC or the U.S. Securities and Exchange Commission - ``SEC'') or introduced by the exchange operators or researchers.
    \item The possible application of such a capability to perform ``stress testing'' of financial market systems (similar in spirit to the Comprehensive Capital Analysis and Review - ``CCAR'' - program introduced for major banks under the auspices of \cite{dodd-frank_2010}).
\end{itemize}

Our research builds upon an extensive modeling effort conducted over the past six years at the Hanlon Financial Systems Center of the Stevens Institute of Technology (Hoboken, NJ - USA), known as the Stevens High Frequency Trading Market Simulation System (SHIFT) project. We aim to demonstrate that this tool is extremely versatile and provides a financial laboratory environment akin to laboratory environments from other research areas - where experiments can be run in isolation, but in realistic conditions. To accomplish this, the rest of this paper is organized as follows. Section \ref{sec:description} provides a description of the system, presenting its modules and some of the design decisions behind them. Section \ref{sec:replay} discusses market event replay capabilities, along with their applications on research and teaching. Section \ref{sec:simulation} presents the use of the platform when creating a completely artificial market through the use of autonomous agents. The agents can reproduce actual market stylized facts, and we study the effects caused by changing their parameters. Section \ref{sec:conclusion} concludes our paper, and presents future possible directions for our work.


%% file: description.tex
\section{System Description}
\label{sec:description}

SHIFT is a complete and standalone system designed to emulate the essential parts of an exchange: a distributed, real-time, and order-driven market. Its initial development focus has been on equity markets, however, the platform can be extended to commodity, future, and option markets, and potentially to any other asset class. The system may be thought of as more of a replica of a real time market exchange rather than a simulation environment.

The platform operates in two different ways. In one mode, SHIFT works with live, real-time, order-level market data sent by market participants which influence everything in the market. This is typically the format used in research studies. This implementation provides researchers the ability to assess unexpected interactions between different strategies. In a second mode, the system replays recorded datasets of quote data. This implementation is typical for commercial market simulators (e.g., paper trading accounts from Interactive Brokers\footnote{\url{https://www.interactivebrokers.com}}, Quantopian\footnote{\url{https://www.quantopian.com}}, etc.) and we normally use this type of implementation for trading competitions and teaching. In either mode, SHIFT is capable of generating trade and quote records that may be used to evaluate the effectiveness of complex trading strategies under conditions similar to a real high frequency market.

A realistic platform such as SHIFT needs to process a massive amount of real-time data, while interacting with an undefined number of clients. Thus one of its major challenges is performance. Particularly, in a high frequency market, speed is a critical factor. To accomplish this, apart from developing in a high performance programming language (C++), we separate the server side of the system in different modules, each with a specialized task. This allow us to avoid overloading any of the modules, as well as to divide the work of each layer of the system into multiple copies of the same module, if necessary.

\begin{figure}[ht]
\centering
\includegraphics[width=5cm]{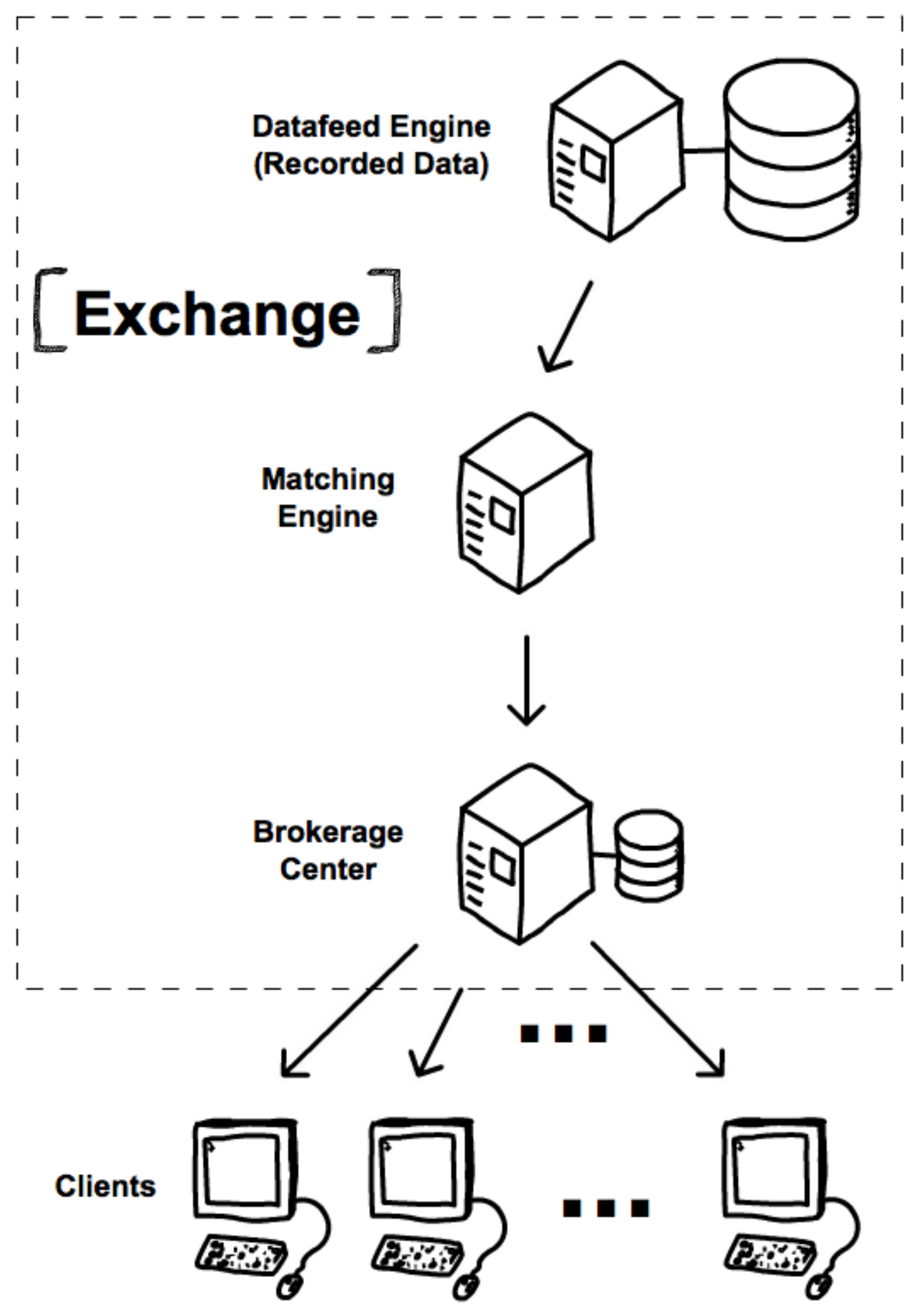}
\caption{Modules of the SHIFT platform.}
\label{fig:shift}
\end{figure}

A simplified schematic of the primary modules in our system is shown in Figure \ref{fig:shift}. The arrows in Figure \ref{fig:shift} point from the server to the client, however information flows both ways in all connected levels of the platform. All communication is done using the Financial Information eXchange (``FIX'') protocol\footnote{We use an open source implementation of FIX 5.0 SP2, QuickFIX. It is available at: \url{http://www.quickfixengine.org}.}, the industry standard.

In the following subsections, we offer a brief description of the system's architecture, with details on both server side, called ``Exchange'', and client side. A note on the scalability of the platform is also given.


\subsection{Exchange}
\label{subsec:exchange}

Our financial market exchange simulator contains three distinct modules: \textbf{Datafeed Engine}; \textbf{Matching Engine}; and \textbf{Brokerage Center}.

\paragraph{Datafeed Engine.} This module works as a streamer of data to the Matching Engine when the system is running in replay mode. It requests and stores historical quoting and trading data from a market data provider, by implementing the necessary API (application programming interface). Replay mode may be used to test single-user trading strategies with historical data, or to provide liquidity in a multi-user environment (e.g., artificial agents or students in a classroom).

\paragraph{Matching Engine.} As it is the case for all its real market counterparts, this module is the brain of our exchange. It is responsible for managing the limit order book (LOB) of all of the platform's traded assets, implementing the dynamics of an order driven market. The matching engine manages a local LOB, containing orders from the clients that are connected to the platform, for each ticker. It also maintains a global LOB, which functions as the National Best Bid and Offer (NBBO) system specific to U.S. markets. The Matching Engine automatically routes orders from the local LOB to the global LOB whenever a better price may be obtained in an outside exchange.

\paragraph{Brokerage Center.} This is the hub that centralizes all communication between clients and Matching Engine. It was initially conceived as a way to remove unnecessary load from the Matching Engine in functions such as providing all current limit order book data to newly connected clients, as well as broadcasting changes in the limit order books to all clients. However, in its current implementation, its use has been expanded. It charges transaction fees (both long and short sells), and it keeps portfolio information for all connected clients, along with their current buying power\footnote{The amount of money a user of the system has available to spend.}. This information is used for account persistence and portfolio valuation, as well as assessing trading limits and margin calls. The Brokerage Center also stores permanent records of all trading data generated by the users of the exchange.


\subsection{Clients}
\label{subsec:clients}

\begin{figure}[ht]
    \centering
    \begin{subfigure}{0.64\textwidth}
        \centering
        \includegraphics[width=\textwidth]{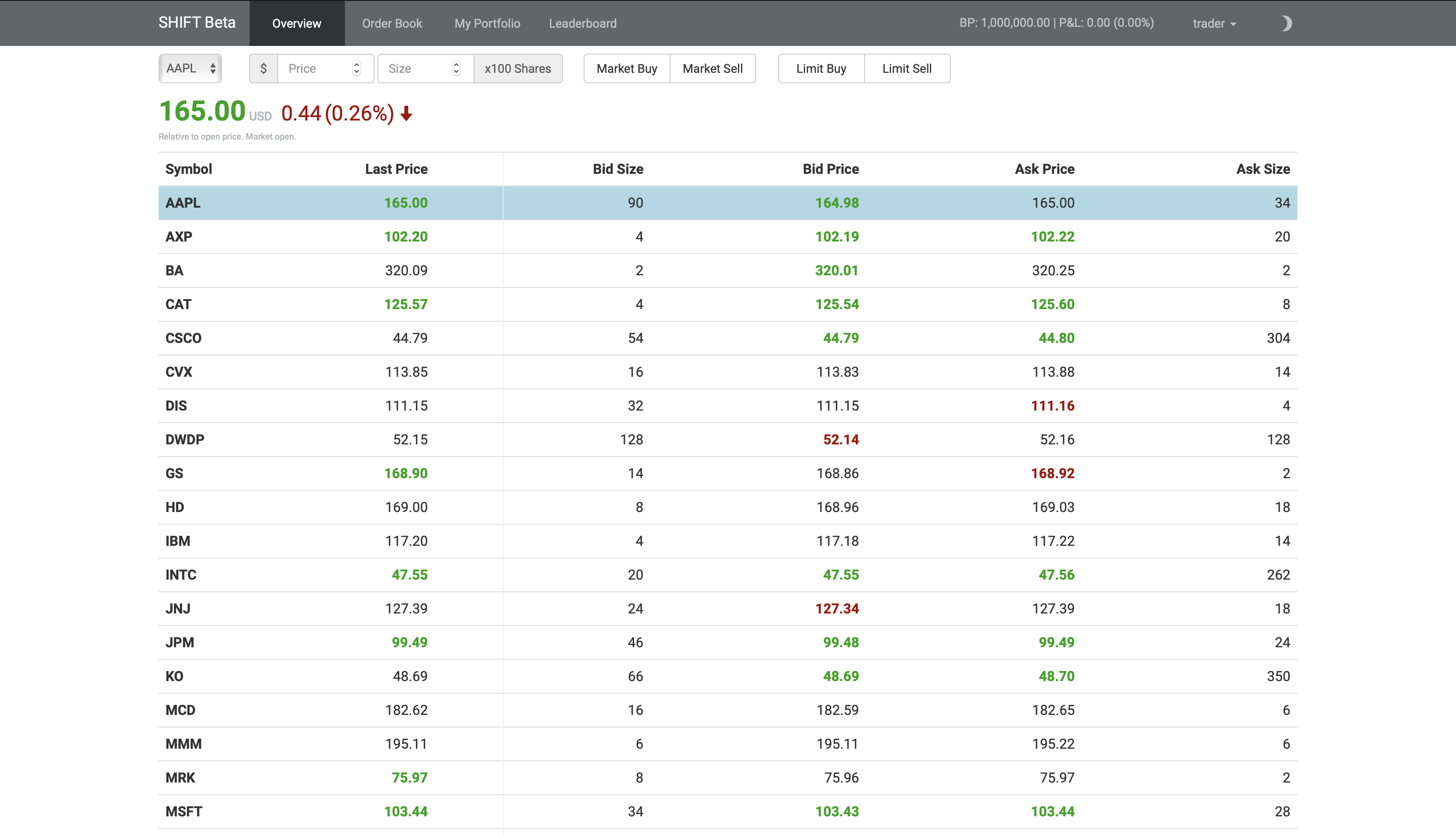}
        \caption{Overview page, with last and best prices data for each of the trading symbols. The green and red coloring indicate up and down movements, respectively, since the last update.}
        \label{fig:wc-overview}
    \end{subfigure}
    \\
    \begin{subfigure}{0.64\textwidth}
        \centering
        \includegraphics[width=\textwidth]{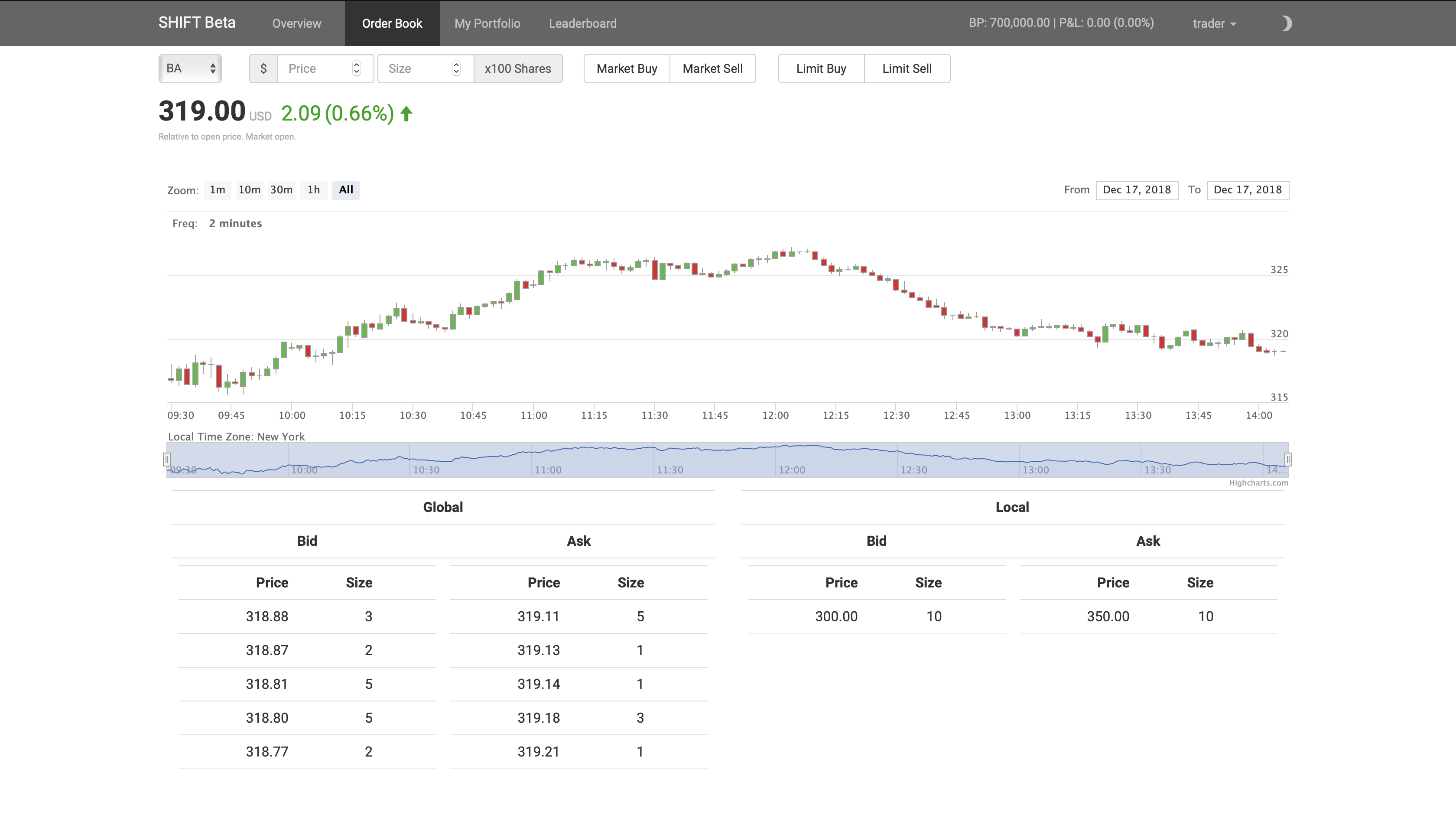}
        \caption{Each trading symbol has its own LOB page, containing a candlestick data plot of the simulated price, as well as the global and local LOBs, as explained in Section \ref{subsec:exchange}.}
        \label{fig:wc-order-book}
    \end{subfigure}
    \caption{SHIFT web interface.}
    \label{fig:webclient}
\end{figure}

We have developed two main ways for users to access our platform: a web interface and APIs in C++ and Python. The web interface was developed with students in mind, so that they could use it in market microstructure classes to learn the rules of operating a trading account in a real market. A sample of the interface is presented in Figure \ref{fig:webclient}. In addition to the overview page (Figure \ref{fig:wc-overview}) and the limit order book page (Figure \ref{fig:wc-order-book}), users can also see their portfolio information.

For more advanced uses, we have created APIs in both C++ and Python. These can be used to create complete algorithmic trading strategies, and we use them in teaching and in research. For research purposes, each \textit{client} can be viewed as an \textit{agent} in an agent-based simulation. Since agents are actual trading accounts operated by individual pieces of software or real people, SHIFT provides a more complex and close-to-reality simulation than existing literature. Because of its server-client architecture, multiple simultaneous agent connections are naturally asynchronous, and even the effects of network latency can be explored.

Examples of use of the platform as an agent-based simulation tool are presented in Section \ref{sec:simulation}. Basic examples of use of our Python API can be found in Appendix \ref{app:api}.


\subsection{Scalability}
\label{subsec:scalability}

\begin{figure}[ht]
    \centering
    \begin{subfigure}{0.47\textwidth}
        \centering
        \includegraphics[width=\textwidth]{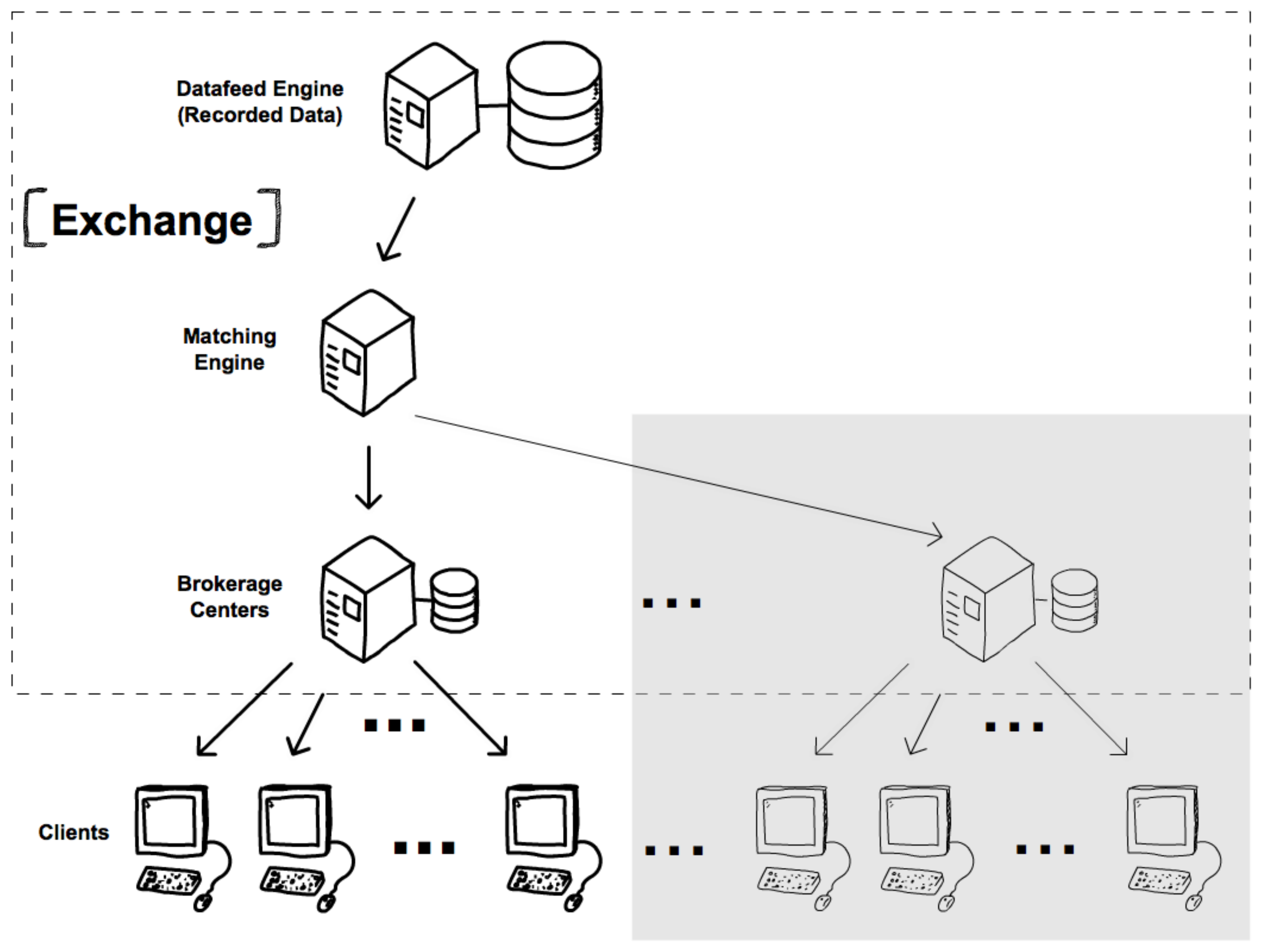}
        \caption{Schematics of the platform with more instances of the Brokerage Center, as a measure to avoid hitting network performance bottlenecks.}
        \label{fig:multiple-bc}
    \end{subfigure}
    \qquad 
    \begin{subfigure}{0.47\textwidth}
        \centering
        \includegraphics[width=\textwidth]{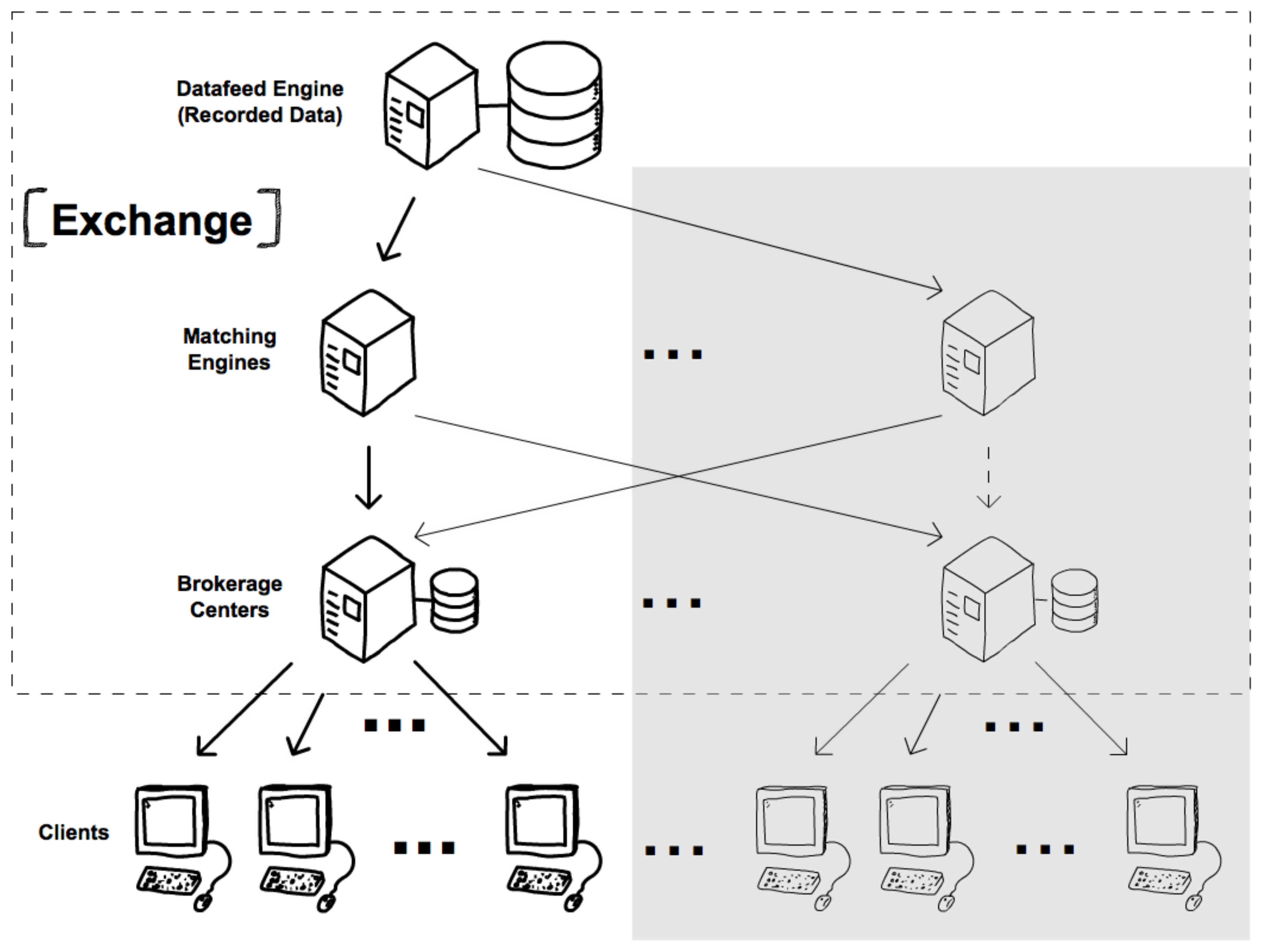}
        \caption{Schematics of the platform with more instances of both the Brokerage Center and the Matching Engine.}
        \label{fig:multiple-bc-me}
    \end{subfigure}
    \caption{Scalability of the SHIFT platform.}
    \label{fig:scalability}
\end{figure}

The platform was developed so that it may be scaled to any number and types of assets as well as any number of clients. The modular architecture allows us to add more instances of each module as needed. For example, a common issue in high frequency studies is when a large number of simultaneous client connections causes the system to slow down due to increased network traffic. A solution is presented in Figure \ref{fig:multiple-bc} where we add more instances of the Brokerage Center.

In the case when the Matching Engine starts receiving more orders than it can process in real time, or if we simply want to add different financial assets, we may add more Matching Engine modules (Figure \ref{fig:multiple-bc-me}).

%% file: replay.tex


\section{Replaying Market Events}
\label{sec:replay}

Outstanding orders of users connected to the platform are placed in what we call the local limit order book. These orders follow the usual rules of order-driven markets, with price-time priority of orders. When in replay mode, the system also makes use of market data obtained from a particular provider. We currently collect microsecond last and best prices data from different exchanges, along with their volume, and we use this information to create what we call the global limit order book of each asset - representing the National Best Bid and Offer (NBBO) system.

The Datafeed Engine streams data to the Matching Engine, which keeps track of the best prices as they were at a given moment in time. These global quotes together with the orders coming from the users in the system create market liquidity. Liquidity therefore is not infinite in the system. There are two major consequences of this design. First, users are in fact competing for liquidity, so two equal orders submitted at exact the same time may have completely different outcomes, depending on which order arrives first at the Matching Engine. Second, even though users cannot cause long-term impact on market prices in replay mode, traded prices may deviate from the real prices for a little while.

In general, researchers and students who backtest trading strategies use downloaded historical price and quotes data. They therefore use unrealistic assumptions such as infinite liquidity and minimal reaction time. In our system we can account for order timing, bid-ask spread, and available volume thus creating much more realistic results. Moreover, the capability of replaying any given day, or of creating completely artificial market scenarios (see Section \ref{sec:simulation}), allows researchers to better design take profit and stop loss rules, as well as stress test their trading strategies.


\subsection{Using the System as a Tool to Understand Market Dynamics}
\label{subsec:tool}

In an effort to engage students with hands-on experience with modeling and algorithmic trading, we introduced SHIFT into lectures at Stevens Institute of Technology. From computing basic statistics from a live stream of limit order book and last price data to implementing and verifying their own trading strategies in market microstructure and algorithmic trading classes, the feedback from students so far has been very positive. We should mention that SHIFT is invaluable in demonstrating to the students a specific point. Every strategy that we implemented which is profitable when using daily data ends up losing money in a realistic system using intraday, high frequency data. 

A pilot algorithmic trading competition ran during the Spring semester of $2019$, and others are planned for the future. In this first edition, there were $38$ participating students, divided in $11$ teams of $3$ to $4$ students each, trading any of the $30$ Dow Jones Industrial Average\footnote{https://us.spindices.com/indices/equity/dow-jones-industrial-average} stocks. Each week, teams were given access to their own instance of the simulator for $6$ days of training, which culminated in all $11$ algorithmic trading strategies running against each other and competing for the best opportunities in day $7$. Every competition day had a different theme, from low volatility days to flash crash days, no outside (human) intervention was allowed, and portfolios would reset, giving a fair chance for teams to recover from a bad week. In the end, the team with the highest total profit after $6$ weeks of competition won.

\begin{figure}[ht]
    \centering
    \includegraphics[width=0.54\textwidth]{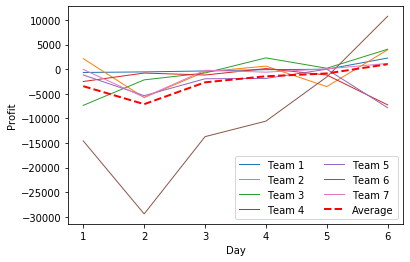}
    \caption{Algorithmic trading competition teams' daily profit evolution.}
    \label{fig:competition}
\end{figure}

The competition was beneficial for us since it allowed us to discover and fix many issues as well as improve the system usability. It was also beneficial for the students who learned about trading and difficulties of applying class concepts to real world. Figure \ref{fig:competition} shows the daily profit of the top $7$ teams along with their average (red dotted line) during the competition's $6$ weeks. The lines generally display a positive trend showing that students were learning from their mistakes and enhancing their algorithms.

%% file: simulation.tex
\section{Artificial Market Capabilities}
\label{sec:simulation}

When not replaying market events, the global limit order book functionality is turned off, and all market formation happens in the local limit order book, with orders coming from the users of the system. These can be researchers, students, market practitioners, or completely artificial agents.

As an initial proof of concept, we set ourselves to create the simplest possible market, where zero intelligence agents with no notion of profit or loss trade a single asset. We describe these agents in the next subsection, followed by the results of experiments we did with such agents in Sections \ref{subsec:replica} and \ref{subsec:stress}.


\subsection{Zero Intelligence Agents}
\label{subsec:zero}

The trading strategy we chose for our zero intelligence agents is inspired from previous work done in the Genoa Artificial Stock Market, described in \cite{raberto_agent-based_2001} and \cite{ponta_statistical_2012}. Modifications were necessary due to the real-time nature of our simulation.


\subsubsection{Trading Strategy}
\label{subsubsec:strategy}

During the trading session (i.e. a simulation execution), each agent trades according to a Poisson process with fixed rate $\lambda$ ($\lambda$ is the same for every trader). One can find details of generating a Poisson process in \cite[Chapter 10]{florescu_probability_2014}. Each of the $N$ traders trades $\Phi_{i}$ times at times  $\tau_{i, j}$, $j = 1...\Phi_{i}$.

At time $\tau_{i, j}$, the $i$-th trader will execute two simple actions:
\begin{enumerate}
    \item If the trader has an outstanding order, i.e. if their last limit order (or a portion of it) is still in the limit order book, send a corresponding cancel for the remaining (buy/sell) order.
    \item Decide whether their next limit order is going to be a buy or a sell (probability $0.5$):
    \begin{itemize}
        \item If the order is a limit buy, the limit price of the order will be $P_{\tau_{i, j}}^{b} \sim N(\mu_{\tau_{i, j}}^{b}, \sigma^{2})$, where $\mu_{\tau_{i, j}}^{b}$ is the smaller value between the current best bid and the last available price. This simulates the fact that buyers want to pay the lowest possible value to acquire assets.
        \item If the order is a limit sell, the limit price of the order will be $P_{\tau_{i, j}}^{a} \sim N(\mu_{\tau_{i, j}}^{a}, \sigma^{2})$, where $\mu_{\tau_{i, j}}^{a}$ is the larger value between the current best offer and the last available price. Sellers want to receive the highest possible value for their assets.
    \end{itemize}
\end{enumerate}

An initial price value $P_{0}$ is given as a parameter to our autonomous agents, representing the close price of the previous day. This value is used as the initial $\mu_{\tau_{i, 0}}^{b}$ and $\mu_{\tau_{i, 0}}^{a}$ values if no other information is available at the moment, i.e. if no other agent submitted limit orders yet. Furthermore, the volume of each submitted limit order is determined as a proportion $r_{\tau_{i, j}}$, which we will call the current \textit{confidence level}, of the buying power (for limit bids) or number of shares (for limit offers) the $i$th trader has available at the moment of order submission.


\subsubsection{Wealth Distribution}
\label{subsubsec:wealth}

The GASM papers that inspired our agents implementation use an equal distribution of buying power and amount of shares among their autonomous agents. We discovered that such homogeneous distribution has an important contribution to the resulting price formation process, as will be shown in Section \ref{subsubsec:parameters}. Therefore, we have opted for a randomized wealth distribution in our experiments.

The initial division of shares $S = (S_{1}, ..., S_{N})$, $i = 1...N$, with $N$ the total number of traders in the simulation, follows a Dirichlet distribution. The probability density function of a Dirichlet distribution has the following form:
\begin{equation*}
f(x_{1}, ..., x_{N}; \alpha_{1}, ..., \alpha_{N}) =
\frac{\Gamma\left(\sum_{i = 1}^{N} \alpha_{i}\right)}{\prod_{i = 1}^{N} \Gamma(\alpha_{i})}
\prod_{i = 1}^{N} x_{i}^{\alpha_{i} - 1} \, ,
\end{equation*}
where $x_{i} \in (0, 1)$, $\sum_{i = 1}^{N} x_{i} = 1$, and $\alpha_{1}, ..., \alpha_{N}$ are the concentration parameters.

A symmetric Dirichlet distribution is a particular case of a Dirichlet distribution with $\alpha_{1} = ... = \alpha_{N} = \alpha$. The probability density function is then simplified to:
\begin{equation*}
f(x_{1}, ..., x_{N}; \alpha) =
\frac{\Gamma(\alpha N)}{\Gamma(\alpha)^{N}}
\prod_{i = 1}^{N} x_{i}^{\alpha - 1} \, ,
\end{equation*}
with $\alpha$ the concentration parameter. The higher the value of $\alpha$, the more homogeneous the distribution of wealth is among traders.

When a trader $i$ is assigned $S_{i}$ initial shares, we also give them an initial buying power (cash) equal to their selling power (shares). That is  $BP_{i} = P_{0} S_{i}$, where $P_0$ is the initial share value.


\subsection{Experiment 1: Establishing a Functioning Market Exchange}
\label{subsec:replica}

In this section we will introduce parameters and agents that create a well functioning exchange. Our goal is to demonstrate that the resulting price process has similar characteristics with a real price process during a normal trading day devoid of any financial events. We also aim to study how the agent parameters affect the behavior of the formed price process.  

There are $2,000,000$ available shares of CS1, a fake stock ticker, with an initial price $P_{0} = \$100.00$. The initial market capitalization of CS1 is thus $\$200$ million. Since traders receive a sum of cash equal to their endowed shares the total initial value of assets in the market (sum of all buying and selling powers) is $\$400,000,000$. The rest of the parameters in this experiment are as follows: 
\begin{itemize}
    \item $N = 200$ traders.
	\item Simulation length $M = 23400$ seconds ($6.5$ hours).
    \item Agents attempt to trade an average of $\lambda = 390$ times during the trading session, i.e. on average, they submit one limit order every minute.
    \item Standard deviation from the best prices $\sigma = \$0.10$.
    \item Confidence level $r_{\tau_{i, j}} \sim U(0.2, 0.6)$, i.e. limit order size is uniformly distributed between $20\%$ and $60\%$ of the total buying or selling power of the agents, depending if it is a limit buy or a limit sell order, respectively.
    \item Wealth concentration parameter $\alpha = N = 200$. With this $\alpha$ value, traders on average have $10,000$ initial shares each, with a standard deviation of about $670$ initial shares among all traders.
\end{itemize}

\begin{figure}[ht]
    \centering
    \begin{subfigure}{0.49\textwidth}
        \centering
        \includegraphics[width=\textwidth]{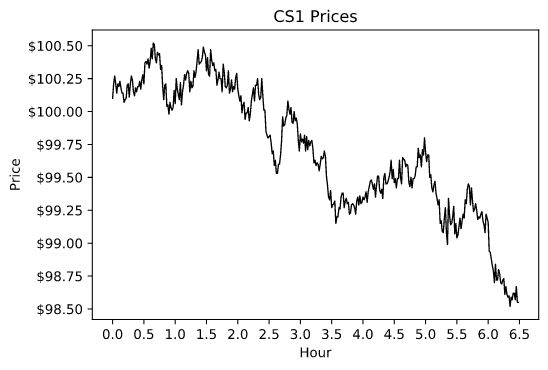}
        \caption{Baseline Experiment 1}
        \label{fig:0010-1t-last_prices}
    \end{subfigure}
    \begin{subfigure}{0.49\textwidth}
        \centering
        \includegraphics[width=\textwidth]{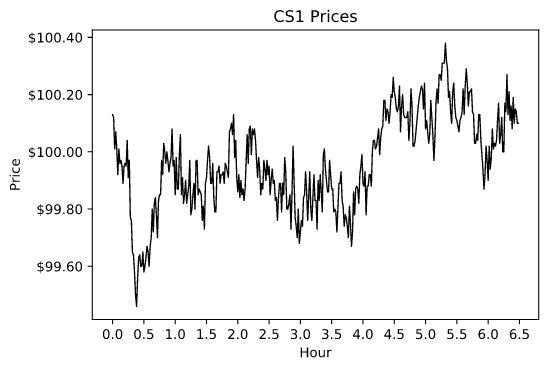}
        \caption{Baseline Experiment 2}
        \label{fig:0015-1t-last_prices}
    \end{subfigure}
    \caption{$1$-minute price paths for two baseline experiments.}
    \label{fig:last_prices}
\end{figure}

\begin{figure}[ht]
    \centering
    \begin{subfigure}{0.49\textwidth}
        \centering
        \includegraphics[width=\textwidth]{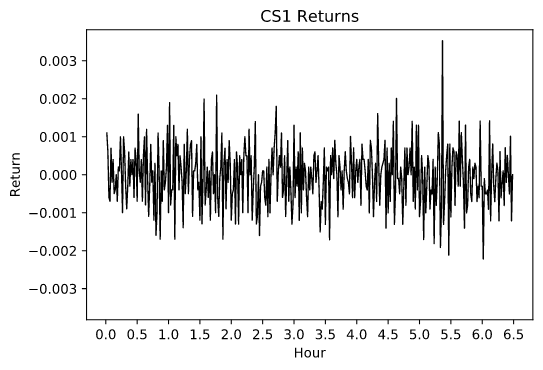}
        \caption{Baseline Experiment 1}
        \label{fig:0010-1t-returns}
    \end{subfigure}
    \begin{subfigure}{0.49\textwidth}
        \centering
        \includegraphics[width=\textwidth]{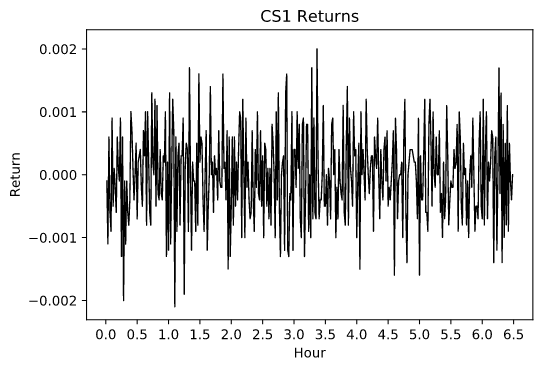}
        \caption{Baseline Experiment 2}
        \label{fig:0015-1t-returns}
    \end{subfigure}
    \caption{$1$-minute returns for two baseline experiments.}
    \label{fig:returns}
\end{figure}

Example simulated price paths are shown in Figure \ref{fig:last_prices}, with the respective return plots in Figure \ref{fig:returns}. Visually, these plots resemble real stock price/return behavior during a given day. Even though there is no flux of outside information into the system, prices display characteristics of real price series, which we discuss in the next sections.

\subsubsection{Comparing Statistics of Simulated and Real Traded Price Data}
\label{subsubsec:statistics}

\begin{figure}[ht]
    \centering
    \begin{subfigure}{0.34\textwidth}
        \centering
        \includegraphics[width=\textwidth]{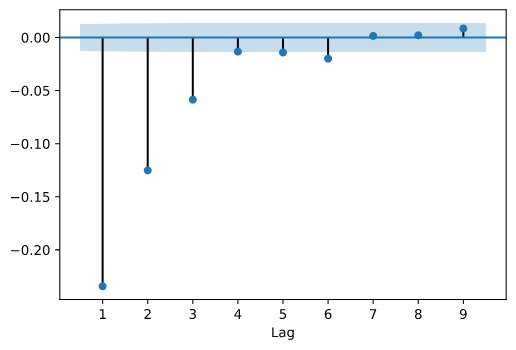}
        \caption{Returns Autocorrelation}
        \label{fig:0010-1s-acf}
    \end{subfigure}
    \begin{subfigure}{0.31\textwidth}
        \centering
        \includegraphics[width=\textwidth]{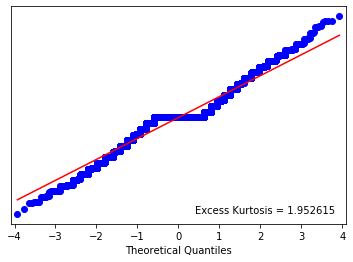}
        \caption{Normal Q-Q Plot}
        \label{fig:0010-1s-qq}
    \end{subfigure}
    \begin{subfigure}{0.34\textwidth}
        \centering
        \includegraphics[width=\textwidth]{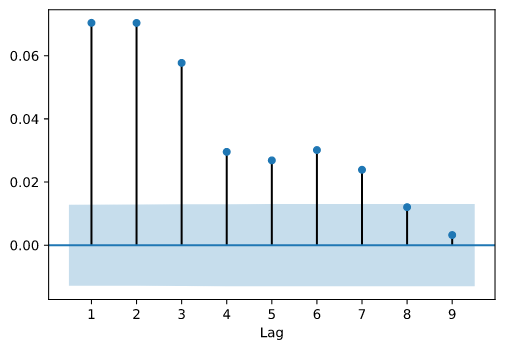}
        \caption{Squared Returns Autocorrelation}
        \label{fig:0010-1s-sacf}
    \end{subfigure}
    \caption{$1$-second returns statistics for a baseline experiment.}
    \label{fig:0010-stylized_facts}
\end{figure}

\begin{figure}[ht]
    \centering
    \begin{subfigure}{0.34\textwidth}
        \centering
        \includegraphics[width=\textwidth]{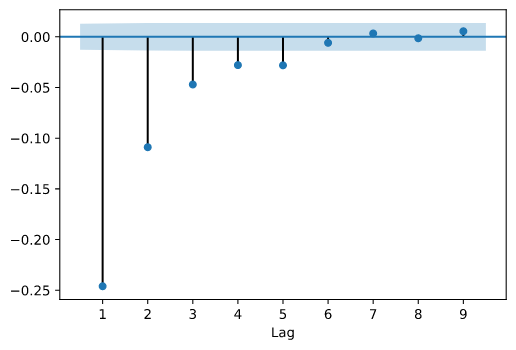}
        \caption{Returns Autocorrelation}
        \label{fig:0015-1s-acf}
    \end{subfigure}
    \begin{subfigure}{0.31\textwidth}
        \centering
        \includegraphics[width=\textwidth]{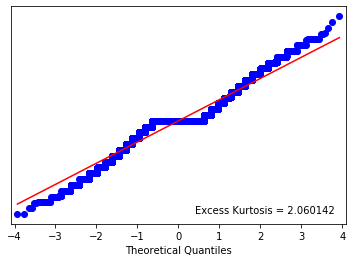}
        \caption{Normal Q-Q Plot}
        \label{fig:0015-1s-qq}
    \end{subfigure}
    \begin{subfigure}{0.34\textwidth}
        \centering
        \includegraphics[width=\textwidth]{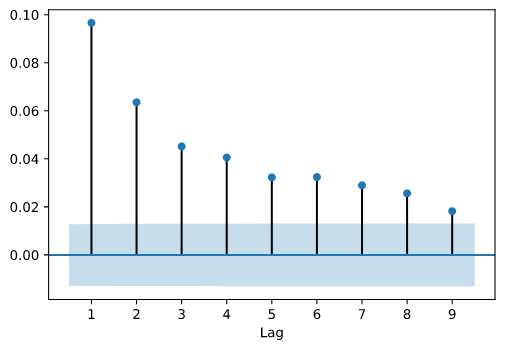}
        \caption{Squared Returns Autocorrelation}
        \label{fig:0015-1s-sacf}
    \end{subfigure}
    \caption{$1$-second returns statistics for a second baseline experiment.}
    \label{fig:0015-stylized_facts}
\end{figure}

Figures \ref{fig:0010-stylized_facts} and \ref{fig:0015-stylized_facts} present statistics for the returns displayed in Figure \ref{fig:returns}. Although here we only discuss the results of two simulated series, the statistics of all simulated experiments are very much in line with the known stylized facts of return time series \citep{cont_empirical_2001}. We present two results to display the consistency of the resulting statistics.

\paragraph{Negative autocorrelation.} Because of the ``bounce effect'' caused by the bid-ask spread, where market orders may match against either side of the book, returns are expected to exhibit a negative autocorrelation when sampling in small time scales, as shown in Figures \ref{fig:0010-1s-acf} and \ref{fig:0015-1s-acf}. 

\paragraph{Leptokurtic behavior.} The distribution of returns has ``heavy tails'' \citep{bouchaud_theory_2003, voit_statistical_2005}. That is, return values far from the average are occurring more frequently than if they should if they followed a Gaussian distribution. This is evidenced in the Q-Q plots (Figures \ref{fig:0010-1s-qq} and \ref{fig:0015-1s-qq}), where the excess kurtosis is also reported. Here we use data sampled every second, and the average excess kurtosis of all experiments we did was around $2$.

\paragraph{Volatility clustering.} When looking at the realized volatility, more precisely at the autocorrelation function of squared returns, it is possible to see in Figures \ref{fig:0010-1s-sacf} and \ref{fig:0015-1s-sacf} that periods of high volatility will lead to other periods of high volatility. This phenomenon, known as volatility clustering, is another known feature of financial market data \citep{teyssiere_volatility_2007}. In fact, the slow decay found in these autocorrelation plots, showing signs of long memory in the volatility, is also documented in the literature \citep{lobato_long_2000}.

Furthermore, the resulting time series data exhibits heteroskedastic effects. When performing the autoregressive conditional heteroscedasticity (ARCH) test \citep{engle_autoregressive_1982} we obtain $p$-values extremely low and we reject the null hypothesis of no ARCH effects for all experiments. If we try to fit an actual ARCH model, we need around $9$ lags to best fit returns sampled every second.

We note that unlike the GASM model our zero intelligence agents do not look at the realized volatility and adapt their strategy depending on its current value. Indeed, the volatility parameter they use when choosing the price of submitted limit orders remains constant during the whole simulation. Our system does not need the agents to adapt to create a price process with all characteristics mentioned in above. We make this observation since it is argued in literature (see e.g., \cite{lux_volatility_1998}) that the arrival of news and the reaction of agents to the news and the market plays a big role in creating such characteristics. In our system we see that even though there is no external news and the agents are very basic we still observe these market characteristics. We thus argue that implementing and respecting the actual trading rules of current financial markets are instrumental to create a proper market simulation.


\subsubsection{Limit Order Book Dynamics for Simulated Versus Real Data}
\label{subsubsec:lobdynamics}

\begin{figure}[ht]
    \centering
    \begin{subfigure}{0.49\textwidth}
        \centering
        \includegraphics[width=\textwidth]{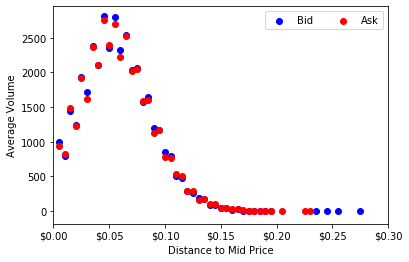}
        \caption{Average Shape of the Limit Order Book}
        \label{fig:0010-1s-lob_shape}
    \end{subfigure}
    \begin{subfigure}{0.49\textwidth}
        \centering
        \includegraphics[width=\textwidth]{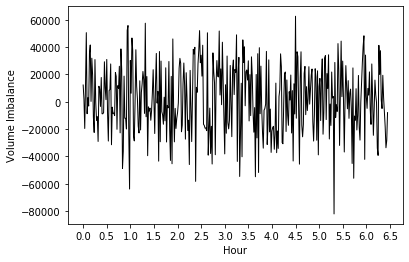}
        \caption{Bid-Ask Volume Imbalance}
        \label{fig:0010-1t-lob_imbalance}
    \end{subfigure}
    \caption{Limit order book average shape and volume imbalance for a baseline experiment.}
    \label{fig:lobdynamics}
\end{figure}

The limit order book data gathered from our simulations displays characteristics found in real market data. Figure \ref{fig:0010-1s-lob_shape} shows the average shape of the limit order book, i.e., the average volume at each tick (in our case, $\$0.01$) distance from the mid price. Here, we present both bids and offers together, since their average volume behavior is the same. In real data this shape is characterized \citep{bouchaud_statistical_2002} by a peak a few ticks away from the mid price, since volumes closer to the mid price tend to be executed more frequently, followed by a power law decay of the average volume of more ``patient'' traders. We observe similar characteristics in Figure \ref{fig:0010-1s-lob_shape}.

\begin{figure}[ht]
    \centering
    \begin{subfigure}{0.34\textwidth}
        \centering
        \includegraphics[width=\textwidth]{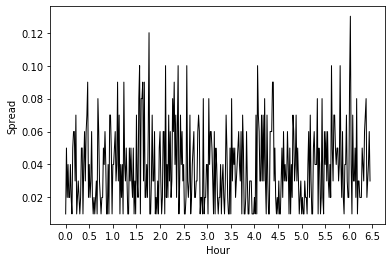}
        \caption{Spread Time Series}
        \label{fig:0010-1t-spread_plot}
    \end{subfigure}
    \begin{subfigure}{0.32\textwidth}
        \centering
        \includegraphics[width=\textwidth]{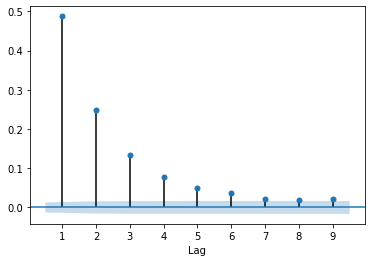}
        \caption{Spread Autocorrelation}
        \label{fig:0010-1s-spread_acf}
    \end{subfigure}
    \begin{subfigure}{0.31\textwidth}
        \centering
        \includegraphics[width=\textwidth]{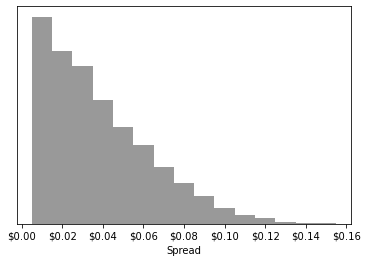}
        \caption{Spread Sample Distribution}
        \label{fig:0010-1s-spread_dist}
    \end{subfigure}
    \caption{Spread characteristics for a baseline experiment.}
    \label{fig:spread}
\end{figure}

In Figure \ref{fig:0010-1t-lob_imbalance}, we plot the dynamic volume imbalance in the limit order book. Specifically, we plot the difference between the volume in each side of the limit order book during the trading day. There are imbalance peaks in both sides of the spectrum throughout the trading day, when there is more pressure from one of the market sides. However, as expected in near-equilibrium, the general trend is mean reverting.

We then turn to spread characteristics in Figure \ref{fig:spread}. Spread is the difference between current best bid and offer prices in the limit order book, and it represents the cost someone incurs when executing a market order. Spread is one of the best proxies for liquidity in high frequency trading \citep{salighehdar_cluster_2017,mago_liquidity_2017}. As previously described in the literature \citep{plerou_quantifying_2005}, the time series of spread values should be characterized by persistence (Figure \ref{fig:0010-1s-spread_acf}). Furthermore, the asymptotic shape of the spread distribution should be described best by a power law (Figure \ref{fig:0010-1s-spread_dist}).


\subsubsection{Connection Between Agent Parameters and Resulting Price Process}
\label{subsubsec:parameters}

Generally agent-based model papers provide a set of parameters that is tested to create market behavior similar to real markets. Here, since the system is so close to reality we can study the impact of parameters on the resulting price formed. Intuitively, in a homogeneous environment we have an entropy/central limit principle that tells us the resulting quantity (temperature, pressure, price) has a Gaussian behavior. The more non-homogeneous the environment the more departure from Gaussianity. Thus we wanted to create parameters characterizing the agents which will allow us to go from homogeneous agents to non-homogeneous ones. 

\paragraph{Impact of order size.} In our base experiment scenario, traders submit buy or sell orders with sizes ranging from $20\%$ to $60\%$ of their current cash or shares value, respectively. This proportion $r_{\tau_{i, j}}$ is random for every trade and represents the trader \textit{confidence level} at the moment when the order is sent. This confidence level turns out to be very important for the distribution of the resulting returns.

\begin{figure}[ht]
    \centering
    \begin{subfigure}{0.31\textwidth}
        \centering
        \includegraphics[width=\textwidth]{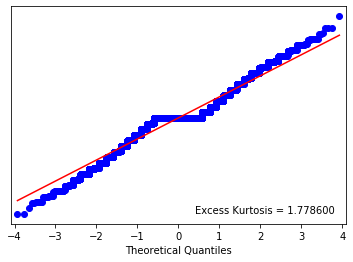}
        \caption{$r_{\tau_{i, j}} \sim U(0.2, 0.4)$}
        \label{fig:w0-cl0}
    \end{subfigure}
    \begin{subfigure}{0.31\textwidth}
        \centering
        \includegraphics[width=\textwidth]{figures/0010-1s-qq.png}
        \caption{$r_{\tau_{i, j}} \sim U(0.2, 0.6)$}
        \label{fig:w0-cl1}
    \end{subfigure}
    \begin{subfigure}{0.31\textwidth}
        \centering
        \includegraphics[width=\textwidth]{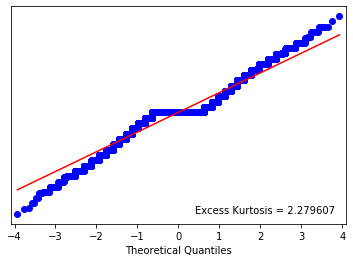}
        \caption{$r_{\tau_{i, j}} \sim U(0.2, 0.8)$}
        \label{fig:w0-cl2}
    \end{subfigure}
    \caption{Normal Q-Q plots for simulated returns with different trader confidence levels.}
    \label{fig:wealth_0}
\end{figure}

We ran three different experiments. We vary the agent confidence levels from conservative ($20\%$ to $40\%$), baseline ($20\%$ to $60\%$), and risky ($20\%$ to $80\%$) and we show the Q-Q plots of the returns in Figure \ref{fig:wealth_0}. We can see from the plots that the larger the orders the traders can execute, the more leptokurtic the resulting return distribution will be. When running several of the same experiment, the average excess kurtosis increases from $1.75$ for the conservative case, $1.96$ for the baseline case, to $2.12$ for the risky case - all of these values being statistically different.

\paragraph{Impact of wealth distribution.} Recall that in the typical agent-based simulations all agents have the same initial wealth. This is a typical homogeneous environment. In SHIFT we wanted to have random initial endowment. This is why we use the Dirichlet distribution. We experimented with modifying the wealth concentration parameter $\alpha$ from $N$ to $1$. When $\alpha = N$ traders have on average $10,000$ initial shares each with a standard deviation of about $670$ shares among all traders. When $\alpha = 1$ the agents are much more heterogeneous from the perspective of initial wealth. Their expected value is still $10,000$, but the standard deviation is now about $9,590$ shares. In our experiments, the trader with the largest amount had $66,500$ shares in the beginning of the trading day, while the poorest trader had only $100$ shares.

\begin{figure}[ht]
    \centering
    \begin{subfigure}{0.31\textwidth}
        \centering
        \includegraphics[width=\textwidth]{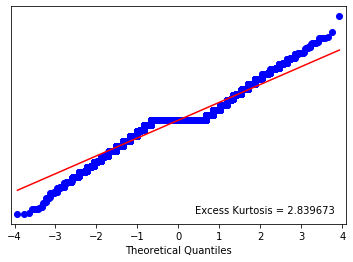}
        \caption{$r_{\tau_{i, j}} \sim U(0.2, 0.4)$}
        \label{fig:w1-cl0}
    \end{subfigure}
    \begin{subfigure}{0.31\textwidth}
        \centering
        \includegraphics[width=\textwidth]{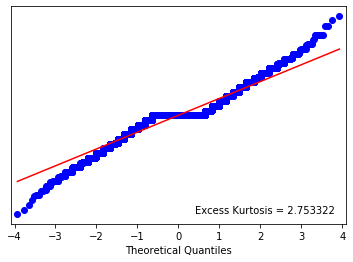}
        \caption{$r_{\tau_{i, j}} \sim U(0.2, 0.6)$}
        \label{fig:w1-cl1}
    \end{subfigure}
    \begin{subfigure}{0.31\textwidth}
        \centering
        \includegraphics[width=\textwidth]{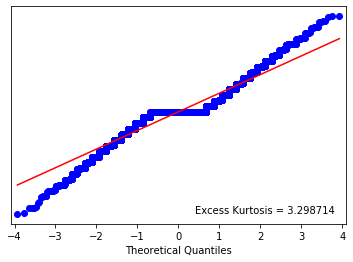}
        \caption{$r_{\tau_{i, j}} \sim U(0.2, 0.8)$}
        \label{fig:w1-cl2}
    \end{subfigure}
    \caption{Normal Q-Q plots for simulated returns with different trader confidence levels, when wealth distribution in non-homogeneous.}
    \label{fig:wealth_1}
\end{figure}

The homogeneous wealth results ($\alpha = N$) have been presented in Figure \ref{fig:wealth_0}. We contrast those results with the results in Figure \ref{fig:wealth_1}. This non-homogeneous distribution of wealth is likely to be closer to reality. Exchanges today have a small number of large institutional traders that dominate through their volume of trades.

The resulting excess kurtosis is always larger than in the previous (more homogeneous) experiments. We note that, although for this particular run Figure \ref{fig:w1-cl1} shows an excess kurtosis below the value in Figure \ref{fig:w1-cl0}, on the average, the relationship between different confidence level ranges stays the same. The average values are $2.80$, $2.98$, and $3.30$ - from conservative to risky cases.

The relationship between ``heavy tails'' and the impact of orders coming from large market participants has been previously studied \citep{gabaix_theory_2003}. It is nonetheless interesting that we can easily reproduce it with simple parameter values changes in our simulation.


\subsubsection{Effects of Different Sampling Frequencies and Trading Activity}
\label{subsubsec:sampling}

We think this section is one of the most interesting observations we made simply by running the system and varying parameters. It is well known that using different sampling frequencies produces different parameter values. For example, in one of the most cited papers in mathematical finance literature \citep{zhang_tale_2005}, the authors observe that realized variance has different values depending on the sampling frequency of the price data used. They attribute this discrepancy to noise in the market and propose a new estimator that is used extensively today (multi-grid realized volatility). However, in our simulations the same exact run produces completely different distribution shapes depending on the sampling frequency used.

\begin{figure}[ht]
    \centering
    \begin{subfigure}{0.31\textwidth}
        \centering
        \includegraphics[width=\textwidth]{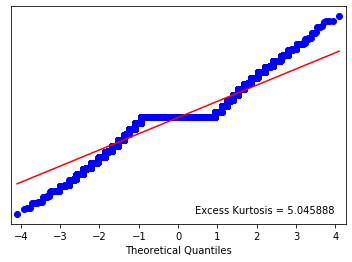}
        \caption{$0.5$-Second Returns}
        \label{fig:_5s_sampling}
    \end{subfigure}
    \begin{subfigure}{0.31\textwidth}
        \centering
        \includegraphics[width=\textwidth]{figures/0010-1s-qq.png}
        \caption{$1$-Second Returns}
        \label{fig:1s_sampling}
    \end{subfigure}
    \begin{subfigure}{0.31\textwidth}
        \centering
        \includegraphics[width=\textwidth]{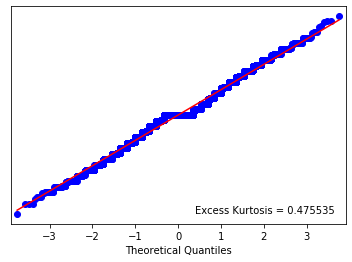}
        \caption{$2$-Second Returns}
        \label{fig:2s_sampling}
    \end{subfigure}
    \caption{Normal Q-Q plots for simulated returns with different sampling frequencies.}
    \label{fig:sampling}
\end{figure}

Figure \ref{fig:sampling} exemplifies such behavior. The returns presented in Figures \ref{fig:_5s_sampling}, \ref{fig:1s_sampling}, and \ref{fig:2s_sampling} are all from the same price series (Figure \ref{fig:0010-1t-last_prices}), but the smaller the time scale, the ``heavier'' the tails of the distribution of returns. This effect is actually present when sampling real financial data as well \citep{aldrich_random_2014}. In fact, this particular effect is called \textit{aggregational Gaussianity} in \cite{cont_empirical_2001}. As the sampling time scale is increased, the returns distribution will get closer to a Normal distribution.

\begin{figure}[ht]
    \centering
    \begin{subfigure}{0.31\textwidth}
        \centering
        \includegraphics[width=\textwidth]{figures/0010-1s-qq.png}
        \caption{$1$-Second Returns when $\lambda = 390$}
        \label{fig:1s-lambda390}
    \end{subfigure}
    \qquad
    \begin{subfigure}{0.31\textwidth}
        \centering
        \includegraphics[width=\textwidth]{figures/0010-2s-qq.png}
        \caption{$2$-Second Returns when $\lambda = 390$}
        \label{fig:2s-lambda390}
    \end{subfigure}
    \\
    \begin{subfigure}{0.31\textwidth}
        \centering
        \includegraphics[width=\textwidth]{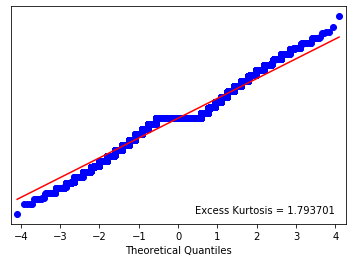}
        \caption{$0.5$-Second Returns when $\lambda = 780$}
        \label{fig:_5s-lambda780}
    \end{subfigure}
    \qquad
    \begin{subfigure}{0.31\textwidth}
        \centering
        \includegraphics[width=\textwidth]{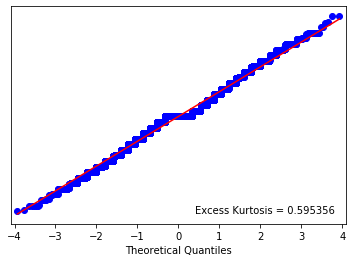}
        \caption{$1$-Second Returns when $\lambda = 780$}
        \label{fig:1s-lambda780}
    \end{subfigure}
    \caption{Normal Q-Q plots for simulated returns. Top row presents results from an experiment where agents submit orders on average every minute ($\lambda = 390$). Bottom row are results from an experiment where traders act on average every half a minute ($\lambda = 780$).}
    \label{fig:lambdas1}
\end{figure}

\begin{figure}[ht]
    \centering
    \begin{subfigure}{0.31\textwidth}
        \centering
        \includegraphics[width=\textwidth]{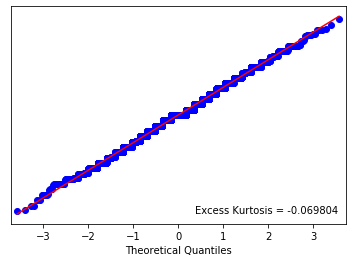}
        \caption{$4$-Second Returns when $\lambda = 390$}
        \label{fig:4s-lambda390}
    \end{subfigure}
    \qquad
    \begin{subfigure}{0.31\textwidth}
        \centering
        \includegraphics[width=\textwidth]{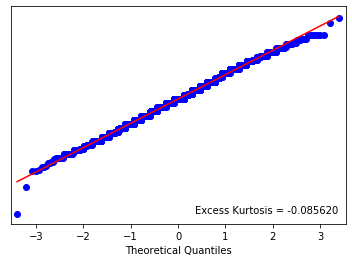}
        \caption{$8$-Second Returns when $\lambda = 390$}
        \label{fig:8s-lambda390}
    \end{subfigure}
    \\
    \begin{subfigure}{0.31\textwidth}
        \centering
        \includegraphics[width=\textwidth]{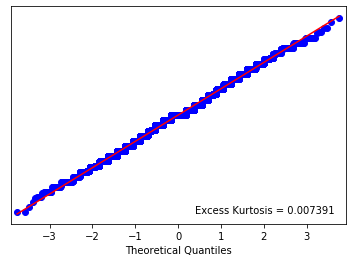}
        \caption{$2$-Second Returns when $\lambda = 780$}
        \label{fig:2s-lambda780}
    \end{subfigure}
    \qquad
    \begin{subfigure}{0.31\textwidth}
        \centering
        \includegraphics[width=\textwidth]{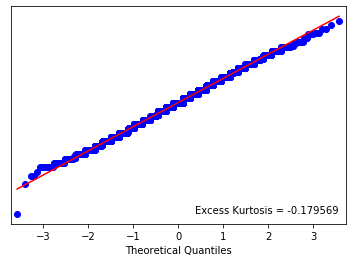}
        \caption{$4$-Second Returns when $\lambda = 780$}
        \label{fig:4s-lambda780}
    \end{subfigure}
    \caption{Normal Q-Q plots for simulated returns. Top row presents results from an experiment where agents submit orders on average every minute ($\lambda = 390$). Bottom row are results from an experiment where traders act on average every half a minute ($\lambda = 780$).}
    \label{fig:lambdas2}
\end{figure}

Moreover, we found that this aggregational Gaussianity is not only related to the sampling time interval, but also to the total trading activity in the market. In the base experiment scenario, with agents submitting orders on average every minute ($\lambda = 390$), we averaged $98,757,500$ shares traded during a simulation day. If we increase their action frequency to once every half a minute ($\lambda = 780$), we averaged $197,323,570$ shares traded during a simulation day. This increase corresponds to a more active equity.

Figures \ref{fig:lambdas1} and \ref{fig:lambdas2} exemplify the relation between sampling frequency, trading activity, and the Gaussian distribution. First, we note the aggregational Gaussianity - specifically we see the excess kurtosis dropping as the sampling interval increases. Second, an even more interesting phenomenon is observed by looking at the ``two'' equities: the baseline and the more active equity. As we double the trading activity and the sampling frequency we obtain similar kurtosis values with the baseline case. Specifically, compare Figure \ref{fig:1s-lambda390} with Figure \ref{fig:_5s-lambda780} and Figure \ref{fig:2s-lambda390} with Figure \ref{fig:1s-lambda780}. We observe a similar phenomenon in Figure \ref{fig:lambdas2}, where we decrease the sampling frequency further. In these results the excess kurtosis values are negligible, but the Q-Q plots visual resemblance is present. This is interesting because it points toward studying and comparing different financial asset time series differently depending on their characteristics, such as trading volume.


\subsection{Experiment 2: Market Stress Scenarios}
\label{subsec:stress}

Following our findings on replicating stylized facts in the context of SHIFT, we demonstrate the system capability to study market stress conditions. Specifically, we study the relationship between market factors and crash characteristics.

To set up the experiments we use $N = 200$ traders, and the simulation length is set to $M = 3600$ seconds ($1$ hour). Around $30$ minutes into the simulation, we create a crash by having new trader(s) forcefully placing a large order on the market. We study the differences in the way we create the crash and the interaction with the market conditions. 

\paragraph{Market factors:}
\begin{itemize}
    \item \textbf{Trading frequency}: Market traders attempt to trade on average every minute (\textit{1 min}) or every half a minute (\textit{0.5 min}).
    \item \textbf{Homogeneity}: The market can be homogeneous (\textit{H}), with an even distribution of wealth ($\alpha = N$) and traders confidence level $r_{\tau_{i, j}} \sim U(0.2, 0.4)$, or non-homogeneous (\textit{NH}), with an uneven distribution of wealth ($\alpha = 1$) and traders confidence level $r_{\tau_{i, j}} \sim U(0.2, 0.8)$. That is, we choose the extreme cases described in Section \ref{subsubsec:parameters} to represent the homogeneous and the heterogeneous market conditions.
\end{itemize}

\paragraph{Crash condition factors:}
\begin{itemize}
    \item \textbf{Stress size}: Crash traders own $5\%$ (level one of the factor) or $10\%$ (level two of the factor) of the total amount of shares available in the market.
    \item \textbf{Stress traders}: This factor has three levels. The first level is a single crash trader placing a large order around $30$ min into the simulation (labeled in the output with \textit{1}). For the second level we consider $20$ crash traders collectively owning the same quantity as the one trader and all placing the orders at about the same time (\textit{20 S} simultaneously). For the third level, we consider $20$ crash traders placing the same total quantity with a $3$-second interval between their actions (\textit{20 NS} non-simultaneously).
\end{itemize}

We ran $10$ experiments for each possible combination of factors, for a total of $240$ experiments. We were initially planning more experiments for each factor combination, but the results were very stable. Sections \ref{subsubsec:drawdown} and \ref{subsubsec:impact} discuss the results obtained.


\subsubsection{Market Drawdown Analysis}
\label{subsubsec:drawdown}

\begin{figure}[ht]
    \centering
    \begin{subfigure}{0.33\textwidth}
        \centering
        \includegraphics[width=\textwidth]{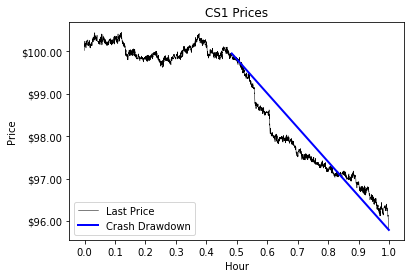}
        \caption{Large Market Impact ($\approx 4\%$)}
        \label{fig:9527-1s-stress_prices}
    \end{subfigure}
    \begin{subfigure}{0.33\textwidth}
        \centering
        \includegraphics[width=\textwidth]{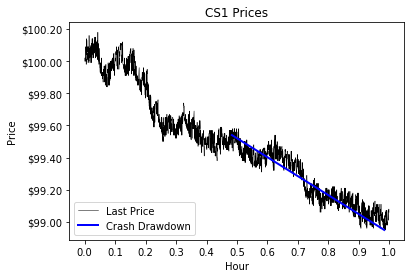}
        \caption{Small Market Impact ($\approx 0.5\%$)}
        \label{fig:6408-1s-stress_prices}
    \end{subfigure}
    \begin{subfigure}{0.33\textwidth}
        \centering
        \includegraphics[width=\textwidth]{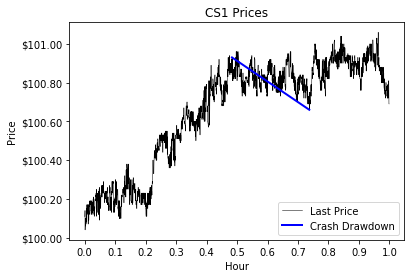}
        \caption{No Market Impact}
        \label{fig:6327-1s-stress_prices}
    \end{subfigure}
    \caption{$1$-second returns during stress events.}
    \label{fig:stress-drawdown}
\end{figure}

In the vast majority of our stress event experiments, the price of the CS1 stock falls after the sell-off event. In some cases, the price drop was considerably large, as exemplified in Figure \ref{fig:9527-1s-stress_prices}. In other cases, the price decrease was not large, and the market would either continue to drop slowly after the stress event (Figures \ref{fig:6408-1s-stress_prices}) or completely recover (Figure \ref{fig:6327-1s-stress_prices}).

Based on the results obtained we analyze which of the factors listed in Section \ref{subsec:stress} are influential for a stress event. One of the main issues is constructing a variable that measures a crash. We could use the drawdown or the time to drawdown to measure the impact of the crash. Here we chose to use the slope of the market drawdown since we know the exact time when the stress event starts. The slope combines both the drawdown size and duration. We illustrate the drawdown slope in blue in Figure \ref{fig:stress-drawdown}.

Since we use categorical variables as inputs and a quantitative variable (drawdown slope) as output, an analysis of variance (ANOVA) is the most appropriate statistical analysis. We display the ANOVA table of the final model that eliminated all non-significant interaction terms in Table \ref{tab:stress-drawdown-anova}. These results indicate a strong influence of each of the four factors individually, as well as that some factor interaction is significant.

\begin{table}[ht]
\caption{Analysis of variance on the slope of the market drawdown in our simulated stress events.}
\label{tab:stress-drawdown-anova}
\centering
\begin{tabular}{ l | r | r | r | r | r }
\toprule
Factor              & DF    & Sum Sq.        & Mean Sq.       & F-Value   & P-Value \\
\midrule
Trading Freq. (TF)  & $1$   & \num{1.30e-09} & \num{1.30e-09} & $96.507$  & \num{<2e-16} \\
Homogeneity (H)     & $1$   & \num{7.14e-11} & \num{7.14e-11} & $5.315$   & $0.022032$ \\
Stress Size (SS)    & $1$   & \num{2.63e-09} & \num{2.63e-09} & $195.614$ & \num{<2e-16} \\
Stress Traders (ST) & $2$   & \num{1.67e-10} & \num{8.34e-11} & $6.211$   & $0.002361$ \\
TF : SS             & $1$   & \num{2.02e-10} & \num{2.02e-10} & $15.02$   & $0.000139$ \\
H : SS              & $1$   & \num{7.20e-11} & \num{7.20e-11} & $5.357$   & $0.021515$ \\
H : ST              & $2$   & \num{9.72e-11} & \num{4.86e-11} & $3.617$   & $0.02841$ \\
Residuals           & $230$ & \num{3.09e-09} & \num{1.34e-11} &           & \\
\bottomrule
\end{tabular}
\end{table}

As expected, the market conditions do not interact, however, the crash agents characteristics interact with market conditions. Since the three way and more interactions are not significant, we next investigated how the combination of factors affects the drawdown slope. We apply a multiple pairwise procedure (Tukey's honestly significant difference - HSD - test) to the resulting significant factors, and we summarize the results in Table \ref{tab:stress-drawdown-tukey}.

\begin{table}[!ht]
\caption{Tukey's HSD test applied to the resulting significant factors of the analysis of variance in Table \ref{tab:stress-drawdown-anova}.}
\label{tab:stress-drawdown-tukey}
\centering
\begin{tabular}{ l | l | r | l | l | r | r }
\toprule
Factor 1       & Factor 2      & Average 1        & Factor 1       & Factor 2      & Average 2        & Diff. P-Value \\
\midrule
Trading Freq.  &               &                  & Trading Freq.  &               &                  &  \\
$0.5$ min      &               & \num{-1.3525e-5} & $1$ min        &               & \num{-8.7907e-6} & $0$ \\
\midrule
Homogeneity    &               &                  & Homogeneity    &               &                  &  \\
NH             &               & \num{-1.1619e-5} & H              &               & \num{-1.0696e-5} & $0.0220322$ \\
\midrule
Stress Size    &               &                  & Stress Size    &               &                  &  \\
$10\%$         &               & \num{-1.4495e-5} & $5\%$          &               & \num{-7.821e-6}  & $0$ \\
\midrule
Stress Traders &               &                  & Stress Traders &               &                  &  \\
$1$            &               & \num{-1.2024e-5} & $20$ NS        &               & \num{-1.1037e-5} & $0.0524403$ \\
$1$            &               & \num{-1.2024e-5} & $20$ S         &               & \num{-1.0413e-5} & $0.0019054$ \\
$20$ NS        &               & \num{-1.1037e-5} & $20$ S         &               & \num{-1.0413e-5} & $0.5071921$ \\
\midrule
Trading Freq.  & Stress Size   &                  & Trading Freq.  & Stress Size   &                  &  \\
$0.5$ min      & $10\%$        & \num{-1.774e-5}  & $1$ min        & $10\%$        & \num{-1.1250e-5} & $0$ \\
$0.5$ min      & $10\%$        & \num{-1.774e-5}  & $0.5$ min      & $5\%$         & \num{-9.3108e-6} & $0$ \\
$0.5$ min      & $10\%$        & \num{-1.774e-5}  & $1$ min        & $5\%$         & \num{-6.3313e-6} & $0$ \\
$1$ min        & $10\%$        & \num{-1.125e-5}  & $0.5$ min      & $5\%$         & \num{-9.3108e-6} & $0.0186553$ \\
$1$ min        & $10\%$        & \num{-1.125e-5}  & $1$ min        & $5\%$         & \num{-6.3313e-6} & $0$ \\
$0.5$ min      & $5\%$         & \num{-9.3108e-6} & $1$ min        & $5\%$         & \num{-6.3313e-6} & $0.000217$ \\
\midrule
Homogeneity    & Stress Size   &                  & Homogeneity    & Stress Size   &                  &  \\
H              & $10\%$        & \num{-1.4551e-5} & NH             & $10\%$        & \num{-1.4439e-5} & $0.9999999$ \\
H              & $10\%$        & \num{-1.4551e-5} & NH             & $5\%$         & \num{-8.7998e-6} & $0$ \\
H              & $10\%$        & \num{-1.4551e-5} & H              & $5\%$         & \num{-6.8423e-6} & $0$ \\
NH             & $10\%$        & \num{-1.4439e-5} & NH             & $5\%$         & \num{-8.7998e-6} & $0$ \\
NH             & $10\%$        & \num{-1.4439e-5} & H              & $5\%$         & \num{-6.8423e-6} & $0$ \\
NH             & $5\%$         & \num{-8.7998e-6} & H              & $5\%$         & \num{-6.8423e-6} & $0.0068273$ \\
\midrule
Homogeneity    & Traders       &                  & Homogeneity    & Traders       &                  &  \\
NH             & $1$           & \num{-1.316e-5}  & NH             & $20$ NS       & \num{-1.1263e-5} & $0.0206907$ \\
NH             & $1$           & \num{-1.316e-5}  & H              & $1$           & \num{-1.0888e-5} & $0.0070141$ \\
NH             & $1$           & \num{-1.316e-5}  & H              & $20$ NS       & \num{-1.081e-5}  & $0.0044496$ \\
NH             & $1$           & \num{-1.316e-5}  & NH             & $20$ S        & \num{-1.0434e-5} & $0.0005655$ \\
NH             & $1$           & \num{-1.316e-5}  & H              & $20$ S        & \num{-1.0391e-5} & $0.0005298$ \\
NH             & $20$ NS       & \num{-1.1263e-5} & H              & $1$           & \num{-1.0888e-5} & $0.9994443$ \\
NH             & $20$ NS       & \num{-1.1263e-5} & H              & $20$ NS       & \num{-1.081e-5}  & $0.9972526$ \\
NH             & $20$ NS       & \num{-1.1263e-5} & NH             & $20$ S        & \num{-1.0434e-5} & $0.9138295$ \\
NH             & $20$ NS       & \num{-1.1263e-5} & H              & $20$ S        & \num{-1.0391e-5} & $0.9082994$ \\
H              & $1$           & \num{-1.0888e-5} & H              & $20$ NS       & \num{-1.081e-5}  & $0.9999945$ \\
H              & $1$           & \num{-1.0888e-5} & NH             & $20$ S        & \num{-1.0434e-5} & $0.9842938$ \\
H              & $1$           & \num{-1.0888e-5} & H              & $20$ S        & \num{-1.0391e-5} & $0.9825362$ \\
H              & $20$ NS       & \num{-1.081e-5}  & NH             & $20$ S        & \num{-1.0434e-5} & $0.9941547$ \\
H              & $20$ NS       & \num{-1.081e-5}  & H              & $20$ S        & \num{-1.0391e-5} & $0.9933012$ \\
NH             & $20$ S        & \num{-1.0434e-5} & H              & $20$ S        & \num{-1.0391e-5} & $1$ \\
\bottomrule
\end{tabular}
\end{table}

Looking at individual factor effects, we see results that we more or less suspected. A more active (\textit{0.5 min}) market exacerbates the drawdown. A non-homogeneous (\textit{NH}) market creates steeper market drawdown movements. Similarly, when traders liquidate a larger market share ($10\%$), this produces a larger slope.

Looking at the stress traders characteristics produces interesting conclusions. Markets in which the stress event is caused by a single trader have a stronger tendency to steeper market drawdown movements than in markets in which the stress event is caused by $20$ traders. In fact, there is no statistical difference when comparing $20$ simultaneous traders (\textit{20 S}) liquidating their shares and $20$ non-simultaneous traders (\textit{20 NS}) liquidating the same amount. It is easy to understand the difference may exist when comparing a single trader with $20$ traders liquidating the same order but over a longer period. It is the difference between absorbing a sudden shock all at once or in smaller doses. The reasoning why the $20$ simultaneous traders behavior is closer to the $20$ non-simultaneous traders rather than the single trader is not that easy. We believe what we are seeing is related to the price-time order priority of order-driven markets, and the fact that there are other $200$ traders in our simulation competing for this priority. That is, the market order of the single stress trader will be executed in its entirety, all at once. The $20$ orders from the $20$ simultaneous traders are programmed to be submitted all at the same time. However, random orders from the other $200$ traders may arrive between these orders, thus sometimes smoothing the stress event effect. This in turn produces statistically different results. We highlight this finding since such impact may be difficult to observe unless using an order-driven and distributed asynchronous market implementation.

Studying interaction terms, trading frequency and stress size show a clear multiplicative behavior. Less active markets (\textit{1 min}) with stress events of smaller magnitude ($5\%$) show smoother drawdown compared with high active markets (\textit{0.5 min} and $10\%$ stress size). When looking at the interaction between homogeneity and stress size, we see a different picture. Liquidating a $10\%$ order impacts the market much stronger than liquidating $5\%$, regardless of market conditions (\textit{H} versus \textit{NH}). Most interestingly, when studying the interaction between market conditions (homogeneity) and stress traders characteristics (\textit{1} versus \textit{20}), it is clear that stress events caused by a single trader in non-homogeneous market conditions produce a steeper drawdown movement.


\subsubsection{Immediate Market Impact Analysis}
\label{subsubsec:impact}

\begin{figure}[ht]
    \centering
    \begin{subfigure}{0.33\textwidth}
        \centering
        \includegraphics[width=\textwidth]{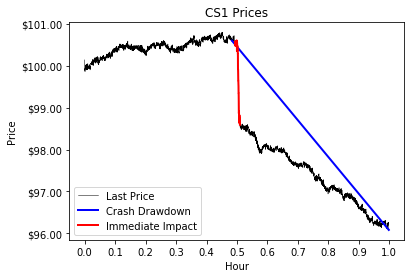}
        \caption{Large Immediate Impact ($\approx 2\%$)}
        \label{fig:9202-1s-stress_prices}
    \end{subfigure}
    \begin{subfigure}{0.33\textwidth}
        \centering
        \includegraphics[width=\textwidth]{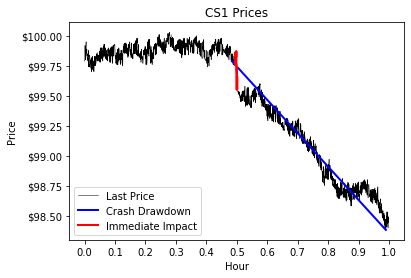}
        \caption{Small Immediate Impact ($\approx 0.2\%$)}
        \label{fig:8003-1s-stress_prices}
    \end{subfigure}
    \begin{subfigure}{0.33\textwidth}
        \centering
        \includegraphics[width=\textwidth]{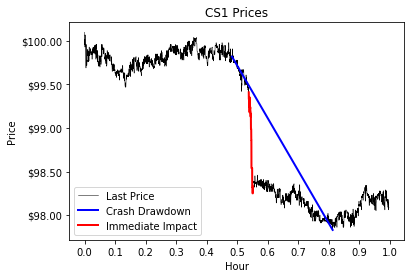}
        \caption{Delayed Immediate Impact ($\approx 1\%$)}
        \label{fig:6129-1s-stress_prices}
    \end{subfigure}
    \caption{$1$-second returns during stress events with immediate impact.}
    \label{fig:stress-impact}
\end{figure}

Drawdown is a classical measure which may be calculated it in our experiments since we know the exact start time of the crash. However, we also want to study the immediate effect of the orders' liquidation on the exchange price. Visually, we can see that some of our experiments show signs of an \textit{immediate impact} in the stock price while others do not (Figure \ref{fig:stress-impact} versus Figure \ref{fig:stress-drawdown}). The price drop may have different magnitudes, as presented in Figures \ref{fig:9202-1s-stress_prices} and \ref{fig:8003-1s-stress_prices}. The drop may in fact happen a few minutes after the sell-off event, as is the case in Figure \ref{fig:6129-1s-stress_prices}.

In order distinguish between the situations in the figures depicted we have to devise a distinguishing criteria. In order to do this we use the return statistics from the identical experiments in Section \ref{subsec:replica} which were lacking the crash traders. Specifically, we use the levels of the two market factors to identify the corresponding non-stress experiments and use its statistics.

The procedure looks at the one second returns of the asset during a window of time starting a few moments before and ending $5$ minutes after the stress event. The largest negative one second return during this period must be greater than $3$ standard deviations from the mean return of the corresponding non-stress simulation day. If there is no such return the market did not experience an immediate impact.

If such a large return exists, we use it as the starting point for further investigation. We denote the time of the largest return with $\tau_{0}$. We next look for returns at least $2$ standard deviations away from the mean return of a calm day, $k$ seconds prior and $k$ seconds past $\tau_{0}$. If they exist, these returns may be positive or negative, since at this point we are interested in any market disturbance that might be part of the immediate impact. We continue to expand our immediate impact window in both directions $k$ seconds at a time until no such returns are found anymore. Finally, the resulting total return (sum of all returns for the period) must be at least $4$ standard deviations away from the mean return of a non-stress simulation day. If everything passes the check the immediate impact period is returned. In practice we use $k = 15$ seconds and $2$ standard deviations as these parameter values maximized the recognition of events with immediate impact visible on the plots, while also minimizing the number of false positives.

This simple technique allow us to identify experiments which had immediate impacts. Next, we analyze which factors had the most influence on the probability of having an immediate impact. The most significant factors and their interactions are presented in Table \ref{tab:stress-impact-pairwise}.

\begin{table}[ht]
\caption{Factors pairwise t-tests for determining which values increase the probability of immediate impact events.}
\label{tab:stress-impact-pairwise}
\centering
\begin{tabular}{ l | l | r | l | l | r | r }
\toprule
Factor 1       & Factor 2    & Average 1 & Factor 1       & Factor 2     & Average 2 & Diff. P-Value \\
\midrule
Homogeneity    &             &           & Homogeneity    &              &           &  \\
NH             &             & $35.0\%$  & H              &              & $45.8\%$  & $0.088$ \\
\midrule
Stress Size    &             &           & Stress Size    &              &           &  \\
$10\%$         &             & $55.0\%$  & $5\%$          &              & $25.8\%$  & \num{2.80e-6} \\
\midrule
Stress Traders &             &           & Stress Traders &             &           &  \\
$1$            &             & $57.5\%$  & $20$ NS        &             & $33.8\%$  & $0.0038$ \\
$1$            &             & $57.5\%$  & $20$ S         &             & $30.0\%$  & $0.001$ \\
$20$ NS        &             & $33.8\%$  & $20$ S         &             & $30.0\%$  & $0.6205$ \\
\midrule
Homogeneity    & Stress Size &           & Homogeneity    & Stress Size &                  &  \\
H              & $10\%$      & $68.3\%$  & NH             & $10\%$      & $41.7\%$  & $0.0072$ \\
H              & $10\%$      & $68.3\%$  & NH             & $5\%$       & $28.3\%$  & \num{1.90e-5} \\
H              & $10\%$      & $68.3\%$  & H              & $5\%$       & $23.3\%$  & \num{1.40e-6} \\
NH             & $10\%$      & $41.7\%$  & NH             & $5\%$       & $28.3\%$  & $0.2314$ \\
NH             & $10\%$      & $41.7\%$  & H              & $5\%$       & $23.3\%$  & $0.0928$ \\
NH             & $5\%$       & $28.3\%$  & H              & $5\%$       & $23.3\%$  & $0.5543$ \\
\bottomrule
\end{tabular}
\end{table}

Both homogeneity and stress size seem to play a large role on the probability of immediate impacts. In fact, about $68\%$ of the simulations in which the market was homogeneous and the stress size was $10\%$ present immediate impact events. The contribution of the stress size on such probability is expected, but the homogeneity behavior is complementing the findings in Table \ref{tab:stress-drawdown-tukey}. In the previous results the non-homogeneous markets created a larger drawdown slope. However, when coupled with results from Table \ref{tab:stress-impact-pairwise} we see that even though heterogeneous markets may suffer more drastically overall from a stress event, the impact is not as sudden and there is a smaller likelihood of an observed crash. 

The conclusions related to the stress traders are the same i.e., the $1$ stress trader creates immediate impact more often than the $20$ traders and there is no significant difference between the $20$-trader cases. Looking at the actual magnitude of the immediate impact, we see that larger stress events ($10\%$) produce an average return drop of $-0.5\%$, as opposed to $-0.3\%$ when the stress event size is $5\%$. The immediate impact is longer if the market is less active, with an average of $43$s against $26$s on more active markets.

The stress traders characteristics affect the immediate impact event duration. When a single trader causes the stress event, the average duration is $29$s, while $20$ simultaneous traders produce an average duration of $34$s. These two quantities are not significantly different. However, when $20$ non-simultaneous traders cause the stress event, the average duration of $45$s is significantly different from the other two cases. This is as expected, as the sell-off in this scenario is executed in waves instead of all at once.


%% file: conclusion.tex
\section{Conclusion}
\label{sec:conclusion}

In 2015, market participants reviewed the Regulation AT \citep{cftc_regulation_2015} proposal. The proposed regulation required that any algorithm needs to be tested in ``laboratory conditions'' before being put into practice. The tool is never explicitly mentioned and the absence of such a tool meant that traders would test algorithms in a replica of a real exchange without any market impact in effect backtesting paper trades. Further, implementing the algorithms in a system accessible to regulators means that proprietary algorithms would be potentially analyzed by regulators.

In fact the CFTC Chairman J. Christopher Giancarlo had the following remarks at FIA Expo Chicago, Illinois, on October 17, 2018:
\begin{quote}
As you know, Regulation AT was an initiative of my predecessor, Chairman Massad. My position was and continues to be that, while there were some good things in the proposal, there were other things that were unacceptable and perhaps unconstitutional, including that proprietary source code used in trading algorithms be accessible without a subpoena at any time to the CFTC and the Justice Department.

At heart, Reg AT is a registration scheme that would put hundreds if not thousands of automated traders under CFTC oversight, a role for which our agency has inadequate resources and capabilities. While I share genuine concerns about the inevitability of some future market disruption exacerbated by automated trading algorithms, there is nothing in Reg AT’s proposed imposition of burdensome fees and registration requirements that will prevent such an event. The blunt act of registering automated traders does not begin to address the complex public policy considerations that arise from the digital revolution in modern markets. Worse is that it would give a false sense of security that the CFTC had regulatorily foreclosed such market disruption, which is impossible. That is why I voted against Reg AT. I do not intend to advance it in its current iteration. \cite{giancarlo_2018}
\end{quote}

This paper details SHIFT, a financial market replica with applications to learning and research. Our goal is to replicate real market conditions rather than create a software specialized in agent-based modeling. SHIFT offers a unique environment combining a \textbf{real pricing mechanism}, a \textbf{distributed asynchronous} market, and \textbf{multi-asset} support. We believe that SHIFT may create an environment where algorithms can be tested and stressed in laboratory conditions. The environment may be setup so that proprietary source code may be tested adequately in absence and without participation of other market participants.

This paper describes the system architecture and discusses several use cases. We show how a simple setup may reproduce known stylized facts of the financial markets such as leptokurtic return distributions and volatility clustering. We investigate the resulting order book dynamics, and show that the system reproduces the known average shape of the order book and statistics of the spread. We hope we convinced the reader that the resulting price process has very similar characteristics to real price behavior.

We think one of the most important contributions of the paper is studying how the price behavior is affected by the trading agents characteristics. We also analyzed a stress experiment in a statistical manner and drew some interesting conclusions. We found that having a single trader with a large order is more likely to produce a market crash than $20$ traders liquidating the same order. However, the impact on the market of the $20$ traders lasts longer and has a larger impact on the price in the long term. A crash event in a non-homogeneous market (market in turmoil) has a larger long term impact but it is less propitious to an immediate impact to the price than a crash in homogeneous market conditions.

Finally, SHIFT has been successfully used in market microstructure classes at Stevens Institute of Technology for over a year and plans for future editions of the algorithmic trading competition are under way. Students have the opportunity to test out what they learn in class by either using the web interface or one of our APIs.

We envision a multitude of experiments to take advantage of our financial laboratory environment. Future work includes analyzing market participants wealth evolution and many potential expansions.


%% file: api.tex
\section{Python API Examples}
\label{app:api}

Here, we present some source code listings to showcase how easy it is to create a simple trading strategy using our Python API\footnote{A full documentation of our API is available at: \url{https://github.com/hanlonlab/shift-python/wiki}.}. Listing \ref{code:connection} exemplifies how to import our library (API), create a trader object, and how to connect and disconnect from the system using the API functions. All the user code is written between the \texttt{connect()} and \texttt{disconnect()} function calls. We show an example of a limit order creation and submission. Orders are created by giving them a type (buy/sell), indicating the ticker traded, the order size, and price.

\begin{lstlisting}[caption=Python API: Connection Example, captionpos=b, float=!ht, frame=single, label=code:connection, language=Python, numbers=left, showstringspaces=false]
import shift
trader = shift.Trader("myusername")
trader.connect("connection.cfg", "mypassword")
...
limit_buy = shift.Order(shift.Order.LIMIT_BUY, "CS1", 1, 100.00)
trader.submit_order(limit_buy)
...
trader.disconnect()
\end{lstlisting}

Listing \ref{code:strategy} contains a very simplistic trading strategy. The strategy buys when price is under $\$95.00$ and sells when price is over $\$105.00$. The \texttt{get\_portfolio\_item()} function is used to get the user's current position in the given instrument. The \texttt{get\_last\_price()} function is used to get the last traded price of the provided symbol. Apart from the price limits, the strategy only maintains one current open position, and it has to have a unit available when shorting. Furthermore, the code will only work for up to $10$ swings between $\$95.00$ and $\$105.00$. Obviously, other more complex strategies may be developed, but we chose to present a simple logic to clearly illustrate the use of our API. The \texttt{Trader} class has a multitude of methods developed in the API to interface with the exchange.

\begin{lstlisting}[caption=Python API: Simple Strategy, captionpos=b, float=!ht, frame=single, label=code:strategy, language=Python, numbers=left, showstringspaces=false]
while num_trades < 10:
    # sleep for 5 seconds
    sleep(5)
    # get current position in CS1
    current_position = trader.get_portfolio_item(symbol).get_shares()
    # get last traded price for CS1
    last_price = trader.get_last_price("CS1")
        
    if current_position == 0 and last_price < 95.00:
        market_buy = shift.Order(shift.Order.MARKET_BUY, "CS1", 1)
        trader.submit_order(market_buy)
    elif current_position > 0 and last_price > 105.00:
        market_sell = shift.Order(shift.Order.MARKET_SELL, "CS1", 1)
        trader.submit_order(market_sell)
        num_trades = num_trades + 1
\end{lstlisting}

In this particular strategy, price is sampled every $5$ seconds. During the $5$ seconds the algorithm sleeps price may have moved beyond the trading limits, possibly missing a trading opportunity. The API also provides the user the ability to observe market changes by setting event handlers. An example is \texttt{on\_last\_price\_updated()} in Listing \ref{code:events}. In this code, a user-defined function will be called every time a price update is received from the exchange. The user can determine which financial instrument had the price update and act accordingly. The API provides other such event handlers.

\begin{lstlisting}[caption=Python API: Simple Strategy with Events, captionpos=b, float=!ht, frame=single, label=code:events, language=Python, numbers=left, showstringspaces=false]
def last_price_updated_event(trader, symbol):
    if symbol == "CS1":
        # get current position in CS1
        current_position = trader.get_portfolio_item(symbol).get_shares()
        # get last traded price for CS1
        last_price = trader.get_last_price("CS1")
        
        if current_position == 0 and last_price < 95.00:
            market_buy = shift.Order(shift.Order.MARKET_BUY, "CS1", 1)
            trader.submit_order(market_buy)
        elif current_position > 0 and last_price > 105.00:
            market_sell = shift.Order(shift.Order.MARKET_SELL, "CS1", 1)
            trader.submit_order(market_sell)

...
trader.connect("connection.cfg", "mypassword")
trader.on_last_price_updated(last_price_updated_event)
...
\end{lstlisting}